 \documentclass[preprint]{aastex}

\shorttitle{Speckle Measures with DSSI. VIII.}
\shortauthors{Horch et al.}

\begin{document}
\newcommand{\beq}{\begin{equation}}
\newcommand{\eeq}{\end{equation}}
\newcommand{\ea}{{\it et al.\ }}


\title{Observations of Binary Stars with the Differential Speckle
Survey Instrument. VIII. Measures of Metal-Poor
Stars and Triple Stars from 2015 to 2018}

\author{Elliott P. Horch\altaffilmark{1,12,13},
Andrei Tokovinin\altaffilmark{2},
Samuel A. Weiss\altaffilmark{1,12},
J\'{a}nos L\"{o}bb\altaffilmark{1},
Dana I. Casetti-Dinescu\altaffilmark{1,14,15},
Nicole M. Granucci\altaffilmark{1},
Nicole M. Hess\altaffilmark{1},
Mark E. Everett\altaffilmark{3},
Gerard T. van Belle\altaffilmark{4,12},
Jennifer G. Winters\altaffilmark{5,12},
Daniel A. Nusdeo\altaffilmark{6,12},
Todd J. Henry\altaffilmark{7,12},
Steve B. Howell\altaffilmark{8,12}, 
Johanna K. Teske\altaffilmark{9,12},
Lea A. Hirsch\altaffilmark{10,12},
Nicholas J. Scott\altaffilmark{8,12}, 
Rachel A. Matson\altaffilmark{8,12}, and
Stephen R. Kane\altaffilmark{11} 
}

\affil{$^{1}$Department of Physics, 
Southern Connecticut State University,
501 Crescent Street, New Haven, CT 06515, USA}

\affil{$^{2}$Cerro Tololo Inter-American Observatory, Casilla 603,
La Serena, Chile}

\affil{$^{3}$National Optical Astronomy Observatory,
950 N. Cherry Ave, Tucson, AZ 85719, USA}

\affil{$^{4}$Lowell Observatory, 1400 West Mars Hill Road, Flagstaff, AZ 86001, USA}

\affil{$^{5}$Harvard-Smithsonian Center for Astrophysics, 60 Garden Street, Cambridge, MA 02138, USA}

\affil{$^{6}$Department of Physics and Astronomy, Georgia State University, 25 Park Place, Atlanta, GA 30302, USA}

\affil{$^{7}$RECONS Institute, Chambersburg, PA 17201, USA}

\affil{$^{8}$NASA Ames Research Center, Moffett Field, CA 94035, USA}

\affil{$^{9}$Carnegie Observatories, 813 Santa Barbara Street, Pasadena, CA 91101, USA}

\affil{$^{10}$Kavli Institute for Particle Astrophysics and Cosmology, Stanford University, Stanford, CA 94305, USA}

\affil{$^{11}$Department of Earth Sciences, University of California, Riverside, CA 92521, USA}

\email{horche2@southernct.edu,
tokovinin@ctio.edu,
weisss4@southernct.edu, janos@lobb.com, danacasetti@gmail.com, granucci.nicole@gmail.com,
hessn1@southernct.edu, 
everett@noao.edu
gerard@lowell.edu,
winters@cfa.harvard.edu,
nusdeo@astro.gsu.edu, thenry@astro.gsu.edu,
steve.b.howell@nasa.gov,
jteske@carnegiescience.edu,
lahirsch@stanford.edu,
nic.scott9@gmail.com, rachel.a.matson@nasa.gov,
skane@ucr.edu}

\altaffiltext{12}{ Visiting Astronomer, Gemini Observatory, which is operated by the Association of Universities for Research in Astronomy, Inc., under a cooperative agreement with the NSF on behalf of the Gemini partnership: the National Science Foundation (United States), the National Research Council (Canada), CONICYT (Chile), Ministerio de Ciencia, Tecnolog\'{i}a e Innovaci\'{o}n Productiva (Argentina), and Minist\'{e}rio da Ci\^{e}ncia, Tecnologia e Inova\c{c}\~{a}o (Brazil). }

\altaffiltext{13}{Adjunct Astronomer, Lowell Observatory, 1400 West Mars Hill Road, Flagstaff, AZ 86001 USA}
\altaffiltext{14}{Astronomical Institute of the Romanian Academy, str. Cutitul de Argint 5, Bucharest, Romania}
\altaffiltext{15}{Radiology and Biomedical Imaging, Yale School of Medicine, 300 Cedar Street, New Haven, 06519, USA}


\begin{abstract}

We present { 248} speckle observations of 43 binary and 19 trinary 
star systems chosen to make progress in
two main areas of investigation: the fundamental properties of metal poor stars and
star formation mechanisms. The observations were taken at the Gemini North and South 
telescopes during the period 2015 July to 2018 April, mainly with the Differential
Speckle Survey Instrument (DSSI), but also with a few early results from the 
new 'Alopeke speckle camera at Gemini North. We find that the astrometry and 
photometry of these observations as a whole are consistent with previous work
at Gemini. We present five new visual orbits for systems important in understanding
metal-poor stars, three of which have orbital periods of less than 4 years, and 
we indicate the degree to which these and future observations
can impact our knowledge of stellar properties and star formation. In particular, 
we find a decrease in mass at fixed spectral type for metal poor stars versus their
solar-metallicity analogues that is consistent with predictions that are made from 
current stellar models.

\end{abstract}


\section{Introduction}

Visual and interferometric binary stars have traditionally been used as tools in 
two main ways in astronomy: (1) to gain insight into the relationship between
fundamental stellar parameters such as mass, radius, luminosity, effective 
temperature, and metallicity, and (2) to provide orbital statistics that are sufficiently robust to
distinguish between predictions of star formation theory. Speckle imaging has
played a key role in both areas over the last four decades because it offers diffraction-limited
resolution in the visible spectrum and is
a very efficient, high-precision technique. Many observations
can be taken per night and the quality of those observations can be very high,
particularly with regard to the astrometric precision, which is of great importance
for determining orbital elements with which to do astrophysics in either of these
lines of investigation.

In recent years, the landscape for speckle imaging has changed significantly. Much
better detectors are now available in the form of electron-multiplying CCD
cameras, which allow not only for fainter sources to be successfully observed, but also 
fainter companions to be detected near a primary star of a given magnitude. This 
realization has paved the way for using the Differential Speckle Survey Instrument (Horch \ea 2009; hereafter, DSSI) at 
Gemini, both for stellar
astrophysics as well as exoplanet science. Our previous work at Gemini
({\it e.g.\ }Horch \ea 2012) demonstrated the ability to see companions approximately
5 magnitudes fainter than the primary at a separation of 0.1 arcseconds under typical
observing conditions.
Another important development is the release of parallax measures
from the {\it Gaia} satellite. Many of the most important ``speckle'' binaries are 
small separation systems, less than 0.2 arcsec. While for systems resolved by {\it Gaia}, the parallax 
measures in the second data release (hereafter, DR2; Gaia Collaboration, 2018) are now often much more
precise than {\it Hipparcos} results (ESA, 1997; van Leeuwen, 2007), the
parallaxes of unresolved binaries { can be affected by the motion of the 
photocenter of the system} and are often not as impressive. It is expected 
that this will be corrected in DR4, but as it stands, even the results in DR2
provide an instant improvement in the 
physical separation and mass information
that can be gained using resolved binary star images for many systems.

We initiated two observing programs highlighting aspects of
binary star science at Gemini soon after 
DSSI was given visiting instrument status there in 2012. These were focused on orbits of metal-poor binary stars
and the occurrence rate of triple stars in nearby systems as a test of star
formation mechanisms. First results on both programs
were published in recent years (Horch \ea 2015a, Tokovinin and Horch, 2016), 
and the current work extends those results in both areas. However, it is interesting to note that
as these programs have continued, they have begun to intertwine. Given the 
sensitivity to large-magnitude difference pairs at Gemini with DSSI, a 
number of the metal-poor
objects have been found to have faint third companions. Recent work of
Moe \ea (submitted 2018), suggests an anti-correlation between close ($P < 10^{4}$ days) 
binary fraction and metallicity, and also an increase in wide triples at lower metallicities, so our
small sample may in fact simply mirror their result if it proves to be true. 
In terms of the trinary program,
sustained observations of the small-separation pairs in these systems
can lead to high-precision orbits, aiding our knowledge of stellar masses
for a range of metallicity. Given this, we detail here the latest results on
both programs together, and report the discovery of 12 previously unknown or
unresolved components.


\section{Observations and Data Reduction}

The observations described here were taken at the Gemini North and South
telescopes over the period of 2015 July through 2018 April. Results from
seven different observing runs have been combined, including two on which 
the new 'Alopeke speckle camera (Scott \ea 2016; Scott \ea in prep)
was used at Gemini North. However, in total,
the observations here represent the use of only about 3 nights of telescope
time, including all calibrations. (Several other approved programs also used DSSI
while it was on the telescope.) This highlights the fact that significant progress can be 
made on these
science goals with a modest investment of observing time, provided that the time is 
consistently scheduled
over a period of years. Figure 1 shows the basic properties of the sample of 
observations presented here; in (a), the magnitude differences are plotted
against the separations determined for these systems. 
Secondary stars as faint as 6 magnitudes fainter than the primary are able to
be successfully observed and measured at separations as close as 0.2 arcseconds. In Figure 1(b), the 
same magnitude differences are plotted versus the system V magnitudes.
The sample is bright ($V=2$ to $11$) and nearby (median distance = 56 pc), 
and Gemini's large aperture enables the
resolution and measurement of extremely small-separation systems. In 
Figure 1(a), a typical 5-$\sigma$ contrast curve is shown from our previous
work at Gemini with DSSI; the fact that the measurements here fill this 
parameter space indicates that our detection capabilities are similar to 
our previous observations. A number of points appear above the curve at extremely
small separations and in fact most of these are below the formal diffraction limit; in these cases we have 
measured the binary parameters using the techniques described in 
Horch \ea (2006) for DSSI data. These systems remain blended in the final reconstructed
images that we derive from the speckle data, 
but nonetheless we find strong agreement in the derived position angles
and separations in the two filters used.
In many cases, these are known to be binary from previous spectroscopic
observations as well.

\subsection{Reduction Method}

The reduction method used here is the same as has been detailed in 
our recent papers (Tokovinin and Horch 2016, Horch \ea 2015a,b, Horch \ea 2012). 
A reconstructed image is made from the raw
data using the technique of bispectral analysis (Lohmann \ea 1983), and 
phase retrieval in the Fourier domain is performed with the relaxation algorithm of
Meng \ea (1990). The reconstructed image in each filter is used to identify
components of the system. Once the position of a component has been 
noted to the nearest pixel, a downhill simplex algorithm is used to fit the power
spectrum of the science target to a cosine squared function in the Fourier
plane. The spacing, orientation, and depth of the fitted fringes are then
converted into a separation, position angle, and magnitude difference for
the system. Before fitting the power spectrum, the science observation must be
deconvolved with the power spectrum of a point source. For most of the observations here, 
the point source used was taken at low zenith distance, and then a dispersion model
was applied that included the altitude and azimuth of the science target at the time of 
observation to derive the power spectrum that was used in the deconvolution. This
makes the point source the best possible match for the dispersion that exists in the
science data. For the remainder of the targets, especially the 'Alopeke observations,
a point source was observed near in time 
and in sky position, and this was used for the deconvolution.

In the case of triple stars, the reduction follows along the same lines, except that 
once the reconstruction is calculated, the rough positions of both components are
noted. These are used to generate a trial power spectrum, which exhibits two 
sets of superimposed fringes, and this is compared with the (deconvolved) 
power spectrum of the data. Then, the downhill simplex method is used to 
minimize the weighted least-squared result between the data and fit, which
in turn yields a final position angle, separation, and magnitude difference value
for both the companion stars.

\subsection{Pixel Scale and Orientation}

The standard method for measuring scale and orientation in speckle imaging is
to mount a slit mask on the telescope and to observe a bright single star. The slits
create fringes on the image plane, which allow the scale to be measured from 
first principles, if the wavelength band of the observation is narrow and the color
of the target star is known. At Gemini, it has not been possible to mount a slit mask
for speckle observations to date, so we instead use a small collection of calibration binaries,
that is, binaries with extremely well-known orbits, often determined with a 
large weighting of data points derived from long baseline optical interferometry 
observations. Table 1 shows the systems used during our Gemini runs for this purpose, 
and the references for the orbits used
to calculate ephemeris positions for the epoch of observations in our case,
where the ephemeris separations ranged from 0.1 to 1.1 arcseconds.  
{ It has been several years since the orbit calculation in a number of those
cases, so we also studied the recent observations listed in the Fourth Interferometric Catalog of 
Measurements of Binary Stars (Hartkopf \ea 2001a; hereafter 4th Interferometric Catalog) to 
identify any current trends in the residuals versus the orbital ephemerides. 
Where they were found, these were incorporated into 
the final scale determination for the run by adding the mean residual onto the ephemeris value 
prior to determining a scale value. The final scale was then a weighted 
average between this value and that obtained when no residuals were incorporated,
with the weighting determined by how clearly the residuals showed that the system
was deviating from the orbit prediction. A similar
procedure was followed for determining the position angle offset, and
our calculations also included the known distortion in the reflective channel of DSSI,
most recently discussed in Horch \ea (2017).}

{ In comparing the scales and orientations derived for the runs shown in Table 1, the 
basic pattern that we have seen at other telescopes emerged here as well, namely that the scale values
fluctuate by as much as 1\% from run to run, and the offset angle relative to celestial
coordinates also varies by over 1$^\circ$. As a consequence, and given that we are combining
data from two different telescopes and two different instruments, we judge that the most
conservative approach is to treat each run independently in terms of the scale determination.
However, this comes at a cost:
it leaves us with only two or three objects with which to determine the scale for each run,
and therefore the uncertainties in the scale are larger than one would ideally like for 
precision astrometry. Once the scale and orientation values have been fixed for each run, we
can compute the average residuals obtained on our scale objects using ephemeris predictions
(that is, not including the trends in residuals of other recent observations discussed above),
and we find that the standard error in the scale determined for each run, when averaged over all runs,
is $0.43 \pm 0.12$\%. For the orientation angle, the result was $0.41 \pm 0.11^{\circ}$. This topic will 
be discussed further in Section 3.2, where comparisons with other known positions of some of
our objects are made. }

{ Looking toward the future, we are actively considering 
how to devise the most precise means for determining pixel scale and orientation. 
For example, when faced
with a similar problem, Tokovinin \ea (2010) used two coherent light sources mounted
at a fixed distance on the telescope spider to create fringes on the image plane; 
this is one possibility. However, the new Gemini
speckle cameras, 'Alopeke (already commissioned at Gemini-North) and Zorro 
(expected to be commissioned in 2019 at Gemini-South), also have the ability to mount
an aperture mask inside the instrument in the collimated beam. This is the most
likely long-term solution to the pixel scale issue for precise astrometry. }

\section{Results}

Table 2 shows the main results of our observations. The columns give:
(1) the Washington Double Star (WDS) number (Mason \ea 2001), which also
gives the right ascension and declination for the object in 2000.0 coordinates;
(2) a secondary identifier, most often
the Henry Draper Catalogue (HD) number for the object;
(3) the Discoverer Designation;
(4) the {\it Hipparcos} Catalogue number (ESA 1997);
(5) the Besselian date of the observation (minus 2000)\footnotemark;
(6) the position angle ($\theta$)
of the
secondary star relative to the primary, with North ($0^{\circ}$) through East ($90^{\circ}$) defining the
positive sense of $\theta$; (7) the separation of the two stars ($\rho$), in
arcseconds; (8) the magnitude difference ($\Delta m$)
of the pair in the filter used; (9) center wavelength of the filter used; and
(10) full width at half maximum of the filter transmission in nanometers.
Position angles have not been precessed from the dates shown and are
left as determined by our analysis procedure, even if inconsistent with
previous measures in the literature. (However, wherever possible, we have noted
these inconsistencies.) 
\footnotetext{ If Julian Dates are preferred, the conversion formula between Bessellian Year and Modified Julian Date is found
at {\tt https://www.usno.navy.mil/USNO/earth-orientation/eo-info/general/date-time-def}.}

Twelve objects have no previous detection
of the companion noted; given that we identified eight doubles in previous
papers in this series containing Gemini speckle observations (Horch \ea 2012 and 
Horch \ea 2015a), 
we suggest discoverer designations of DSG (DSSI-Gemini) 
9 through 20, and will refer to them as such throughout the rest of this
paper. It is of course possible that some of these
discoveries are line-of-site companions; this statistically more likely to be the case for those components 
that have larger separations and magnitude differences as shown both in Horch \ea (2014)
and more recently in Matson \ea (2018).

Figure 2 shows the discovery images for two systems, one new discovery 
and one triple.
In Figures 2(a) and (b), the 692-nm and 880-nm reconstructed images show 
a very faint companion detected at a separation of 1.3 arcseconds from the
primary star of HIP 4754 = BD+15 150. 
In Figures 2(c) and (d), we show reconstructed images of HDS 2092, a triple star where we 
clearly resolve the secondary
into a small-separation pair, recently identified by Tokovinin \ea (2018). 
\subsection{Comments on Newly Resolved Pairs}

Some further brief commentary on the twelve stars that host the previously unknown companions is
given below. Basic observational data discussed for each system is taken from SIMBAD\footnotemark
\hspace{1pt} and the Ninth Catalogue of Spectroscopic Orbits of Binary Stars (Pourbaix \ea 2004, 
hereafter, { 9th} Spectroscopic Catalogue), and the Geneva-Copenhagen spectroscopic survey
(Holmberg \ea 2009).

\footnotetext{\tt http://simbad.u-strasbg.fr/simbad/}

\noindent
$\bullet$ DSG 9 (HIP 4754). This F7 subdwarf is probably a triple system; the faint component
seen in our observations is at much too large a separation to correspond to the spectroscopic
system that has period 347 days as determined by Latham \ea (2002). In 
that same work, the authors determine the metal abundance to be [Fe/H] = $-1.5$. (This
system is shown in Figure 2.)

\noindent
$\bullet$ DSG 10 (HIP 19915). In the Geneva-Copenhagen spectroscopic survey, this system
was found to have metal abundance [Fe/H] of $-$2.01, and it was also noted as a double-lined spectroscopic
binary with mass ratio of $0.904 \pm 0.019$. We see not only what is most likely the spectroscopic component
and previously given discoverer designation YSC 129, but also a faint, wide component at a separation of half an arcsecond.
This system could play an important role in our understanding of the fundamental properties of true Population II stars,
if further observations can be obtained and the orbital motion studied.

\noindent
$\bullet$ DSG 11 (HIP 34105, HR 2683). This variable A star was originally selected as a point source
calibration object for one of the science targets on our program. The {\it Gaia} DR2 parallax puts the 
distance to the system at approximately 86 pc, leading to a projected separation of 5 AU for the binary at present,
if the pair is gravitationally bound.

\noindent
$\bullet$ DSG 12 (HIP 56851). This system was listed as a double-lined spectroscopic binary in the 
Geneva-Copenhagen survey, with mass ratio of 0.983. The component in Table 2, however, has 
a magnitude difference of over 2. Given the parallax of $6.8384 \pm 0.0358$ mas 
and our separation of 0.03 arcseconds, a projected separation of 4.4 AU is obtained for this late-G, metal-poor
sub-giant. Thus, it is likely to be a triple system.

\noindent
$\bullet$ DSG 13 (HIP 57421). Our observations confirm that there are two companions within 0.5 arcseconds
of this previously known speckle double, but it is not an obviously heirarchical system. It is also known to be a double-lined binary
of near equal masses, but with projected separations of 10s of AU for both of the resolved components (if bound), neither
is likely to be the spectroscopic component. Given a large proper motion for the primary star ($\mu_{x} = -85.951$ mas), it
should be straightforward to establish if one or both of these components is a line-of-sight 
companion in the coming years, if further observations are obtained.

\noindent
$\bullet$ DSG 14 (HIP 58669). There are double-lined spectroscopic orbits for both the small-separation and large-separation 
pairs in this triple system (Tokovinin \ea 2015). We resolve the small-separation component for the first time (the wider pair being 
WDS 12018$-$3439 = I 215). 
The primary star is a solar near-twin, with the same metal abundance and early-G spectral type.

\noindent
$\bullet$ DSG 15 (HIP 61158, HR 4771). The small-separation component to this bright, late-A giant is approximately
3 magnitudes fainter than the primary and less than 0.1 arcseconds away, illustrating the power of speckle imaging
at Gemini. Virtually no difference in color between the two stars is noted, so the secondary may be a late-A dwarf.

\noindent
$\bullet$ DSG 16Aa,Ab (HIP 68587). We detect here a very small-separation companion to the brighter star in the known 
pair B 263. The primary would appear to be a near-solar analogue, with [Fe/H] of $-0.02$ and spectral type G3V.


\noindent
$\bullet$ DSG 17 (HIP 72622, HR 5531). Although there is no indication of duplicity for this star in the literature,
we find a small-separation companion to this nearby mid-A sub-giant. Within a few years, orbital motion should be
apparent if the system is gravitationally bound.

\noindent
$\bullet$ DSG 18 (HIP 76424, HR 5804). This star lies 52 pc from the Solar System, but the component we find is at a separation
of 1 arcsecond. Thus, if physically bound to the F3V primary star, the orbital period of the secondary is likely to be in the range of hundreds of years.

\noindent
$\bullet$ DSG 19 (HIP 80925). This system has a single-lined spectroscopic orbit in the 9th Spectroscopic Catalogue 
due to Bopp \ea (1970), but more recently appeared in the list of double-lined systems for which Holmberg \ea (2009)
gave a mass ratio; they obtained a value of $0.919 \pm 0.020$. They also measured the metal
abundance as [Fe/H]=$-0.60$ for this K1V star. { A component with a $\sim$4 arcsecond separation  was discovered
by {\it Hipparcos} (HDS 2335), but that would not be present in the small (2.8$\times$2.8 arcsecond) field of view 
for DSSI at Gemini, and in any case has a much larger magnitude difference than what we detect here. }
We conclude that the component detected here is the spectroscopic one, particularly given that the system is only 22 pc 
from the Solar System and the projected separation is therefore approximately 0.2 AU.

\noindent
$\bullet$ DSG 20 (HIP 81170). We measure the separation of this known double-lined spectroscopic binary for the first
time. The system is very metal poor, with [Fe/H]=$-1.39$, but relatively nearby, at a distance of 43 pc. If further observations
can be obtained in the future at Gemini, the possibility of a spectroscopic-visual orbit exists within the next few years, 
and the system could give information relevant to studies of low-metallicity stars, as discussed in Section 4.2.

\subsection{Relative Astrometry and Photometry}

We can judge the precision of our astrometric results by examining the differences in position
angles and separations obtained on the same observations in the two different filters used. 
These are shown in Figure 3 in two ways. In Figures (a) and (b), these differences
are plotted as a function of average measured separation. As expected, these show that 
at smaller separations, the position angle precision deteriorates, although the scatter remains 
fairly symmetric about zero, indicating no systematic offsets between the two channels
of the instrument. On the other hand, the 
scatter of the separation differences remains fairly constant with separation (again with near-zero
mean). In Figures (c)
and (d), the same differences are plotted as a function of the measured magnitude
difference, whereupon it is clear that in both position angle and separation, the scatter grows as
the magnitude differences increase. Overall, 
for the complete sample, the mean difference is $-0.44 \pm 0.27$ mas in separation
and $+1.19 \pm 0.49$ degrees in position angle; however, if separations are confined to the range
0.05 arcsec to 1.0 arcsec, then these numbers become $-0.37 \pm 0.30$ mas in separation and 
$0.25 \pm 0.14$ degrees in position angle. The standard deviation of the residuals in the
former case is $2.54 \pm 0.21$ mas for separation and $1.21 \pm 0.10$ degrees for position angle.
These numbers imply that for an individual measure (not a difference between two measures of the
same uncertainty, as these numbers represent), the 
precision would be estimated as the standard deviation divided by $\sqrt{2}$, so that, on average, the 
uncertainties in the position angle and separation values given in Table 2 are 1.8 mas and 0.9$^{\circ}$,
respectively. A further improvement could be obtained if the measures in each channel were
averaged; in that case the uncertainties would be reduced by another factor of $\sqrt{2}$, 
resulting in 1.3 mas and 0.6$^{\circ}$. These numbers are very close to those of our previous
studies involving Gemini DSSI data ({\it e.g.\ }0.9 mas and $1.0^{\circ}$ were obtained in Horch \ea 2015a
for a data set with fewer large magnitude difference systems and generally smaller separations).

{ Below the diffraction limit, the measurement uncertainty increases.
For example, the collection of measures with separations below 22 mas (the diffraction limit for the 692-nm
filter) in Table 2 has a standard deviation of
separation differences between the two channels of $3.53 \pm 0.54$ mas, approximately 40\% higher
than the result for all observations with separations between 0.05 and 1 arcsec. Assuming that the linear measurement 
uncertainty is the same in the direction along the position angle coordinate as in the radial direction, then the uncertainty in 
position angle should be related to the separation and the uncertainty in separation by:

\beq
\delta \theta = \arctan [\frac{\delta \rho}{\rho}].
\eeq

\noindent
For the separations below the diffraction limit at Gemini, $\delta \theta$ increases substantially due to the combined 
effects of the dependence on $\rho$ and the larger value of $\delta \rho$ itself,
indicating that larger differences in position angle between the two channels will be seen in this 
separation regime.
For the measures below 22 mas, we find the standard deviation in position angle differences to
be  $11.6 \pm 1.8^{\circ}$, which is roughly consistent with the application of the above formula
using $\delta \rho = 3.53$ mas and the average separation for the measures in question, 14.1 mas.
In that case, the we obtain is 14.1$^{\circ}$.
}

In Figure 4, we investigate the accuracy of the measures, as judged by comparisons with the 
ephemeris predictions for systems with well-known orbits (excluding, of course, systems that were used in the
determination of the scale and orientation). We confine our comparison to 
those systems with either Grade 1 or Grade 2 orbits in the {\it Sixth Catalog of Visual Orbits of 
Binary Stars} (hereafter the 6th Orbit Catalog, Hartkopf \ea 2001b). We further impose the criterion that the orbit
used must list uncertainties to the orbital elements, so that we can propagate those 
uncertainties into the predicted separation and position angle for the epoch in which we 
observed the star. Ephemeris positions and residuals for these objects are shown in Table 3, 
where the astrometry from both channels of DSSI has been averaged before making the comparison.
The columns give (1-3) the WDS, discoverer designation, and {\it Hipparcos} number, (4) the 
date of the observation, (5-6) the position angle and separation derived from the orbital elements for 
the date in question, (7-8) the residuals in each coordinate when comparing to the values in Table 2,
and (9) the orbit grade and reference.
The table and Figure 4 indicate that, to the ability that we can tell, there are no systematic
differences between our measures and the ephemeris predictions. Formally, the mean residual
is 
$-0.50 \pm 1.23$ mas in separation and $0.42 \pm 0.47$ degrees in position angle. 

A final astrometric check was done using four objects with separations above 1 arcsecond
in Table 2 that appear as resolved objects in the {\it Gaia} DR2. 
From the {\it Gaia} positions, we deduce the position angle and relative separations of each pair, and form
the difference in each quantity relative to the values in Table 2. 
The results are shown in Table 4,
where the columns give (1) the WDS designation, (2) the discoverer designation, (3) the Hipparcos number, 
(4) the date of { our} observation, (5,6) the position angle and separation of the pair derived from positions
in DR2, and (7,8) the (speckle minus {\it Gaia}) difference in these quantities. As in Table 3,
we have averaged the results in the two filters shown in Table 2 before using the astrometry in Table 4.
For the ensemble of measures
on these four objects, the mean residual 
in separation is $-1.6 \pm 4.8$ mas and in position angle we obtain $+0.63 \pm 1.08^{\circ}$, 
indicating no evidence of scale or orientation offsets. The scatter is larger than the differences between
channels and the orbit residuals discussed above, but that can be explained by a small amount of 
relative motion of these pairs in between our measures and those of {\it Gaia}, which were not taken
at the same epoch, as well as the fact that three of the four systems are triple stars where the small separation
component is not seen in DR2 but is resolved in our observation, thus potentially affecting the astrometry.

{ The absolute value of the mean residual in separation divided by the average 
separation of the sample yields a value of $0.23 \pm  0.57$\% for the orbit study and $0.13 \pm 0.38$\%
for the {\it Gaia} comparison. If we assume this represents the implied scale offset, then both numbers
are somewhat smaller than (but certainly consistent with) the number quoted for the scale uncertainty
in Section 2.2. The position angle
residuals in both studies are likewise similar to the number given for the position angle uncertainty.
Therefore, we conclude that the values given in Section 2.2 are reasonable estimates of the contributions
to the overall uncertainty in our measures coming from scale and orientation calibration.
To obtain the total uncertainty in separation for any measure in Table 2, one would multiply the
measure by the scale uncertainty and 
add that in quadrature with the internal precision estimate of 1.8 mas as follows:

\beq
\delta \rho = \sqrt{1.8^2 + (\rho \cdot 0.0043)^2},
\eeq

\noindent
where $\rho$ is the observed separation in mas. This yields, for example, $\delta \rho = 1.8$ mas for $\rho = 0.1$ arcsec,
2.7 mas for 0.5 arcsec, and 4.4 mas for 1.0 arcsec.
The total uncertainty in position angle would be obtained by adding the 
internal precision and the offset uncertainty in quadrature; that results in a value of $\sqrt{0.9^2 + 0.4^2} = 1.0^{\circ}$.}

Turning now to the relative photometry, we have studied the magnitude differences derived from 
our observations together with the {\it Hipparcos} values in the few cases where relative photometry
exists in the {\it Hipparcos} Catalogue (ESA 1997). A comparison is shown in Figure 5, where the relationship
between the space-based values and those here appears to be linear to the extent we can judge. 
All our observations that are relevant, regardless of the filter used, are plotted in the figure.
While it is hard to conclude much from so few points and the comparison itself is rough due to the difference
in center wavelength of the filters being compared, we can observe that the scatter of the points 
appears to be roughly in line with previous studies using DSSI (Horch \ea 2012, Horch \ea 2015a). 
It is known from our previous work that the uncertainties in magnitude difference grow  below 
$\Delta m = 0.5$ and above $\Delta m = 3.0$ (Horch \ea 2004);
for data reported here in the range of $0.5 \le \Delta m \le 3$ 
we obtain a standard 
deviation in the $\Delta m$ residual of $0.16 \pm 0.07$ for the three highest precision {\it Hipparcos} values.
This includes the canonical uncertainty
in $\Delta m$ that we have stated in other papers of about 0.1 magnitudes for most observations under
good observing conditions.

 In other papers in this series
(e.g.\ Horch \ea 2015b) we have discussed how the separation of a binary system as well as seeing 
affect the relative photometry because they determine the degree to which the speckle pattern of the 
secondary star is within the isoplanatic patch. However, in the present case, due to the combination
of small separation and excellent seeing, nearly all measures meet the quality criteria we have
set forth in the past for reporting magnitude differences. Nonetheless, we have applied the same
methodology to the current data set and in a small number of cases (of separations larger than 
approximately 1 arcsecond) we list the magnitude difference obtained in the fit as an upper limit,
to indicate that the observation could be affected by speckle decorrelation.

Another situation that we have studied in some detail since our last publication of Gemini data
is that of small-separation, large magnitude difference systems. In Horch \ea (2015a), we presented
observations of the triple system HIP 103987 and derived a visual orbit from the data of the
small-separation component that we now
believe to be spurious. It has since been superseded by the orbit of Tokovinin and Latham (2017),
where those authors warned of possible systematic error in the determination of our astrometry and
photometry of the small-separation pair in this system, DSG 6Aa,Ab. We have reviewed the
data reduction of those observations and at this time retract the measures taken at Besselian 
Years 2012.5739 and 2014.5645 as most likely dominated by atmospheric dispersion. On the other hand, 
our observations
made in 2013 and 2015 appear to be on more firm ground, taken at modest zenith angles and
in rough agreement with the Tokovinin and Latham orbit. Thus, we cannot retract the 2013
observation on the same basis, but it should be viewed as uncertain in the context of this discussion, and we will
not report the only new measure (from 2015) here. We will seek further observations of the system to either confirm
or refute its detection.

\section{New Orbital Elements}

The measures in Table 2, together with those found in the 4th Interferometric Catalog 
allow us to give updated visual 
orbital elements for four spectroscopic binaries and to report orbital elements of one system 
for the first time. These are presented in Table 5. For four of the five systems, namely 
WCK 1Aa,Ab (WDS 01083 + 5455, HIP 5336),
YSC 129 (WDS 06416+3556, HIP 32040),
DSG 7Aa,Ab (WDS 16329+0315, HIP 81023), and
DSG 8 (WDS 23485+2539, HIP 117415), 
a visual-spectroscopic orbit calculation was performed, 
where radial velocities found in the 9th Spectroscopic Catalogue were used. In these
cases, the method was that of Tokovinin (2016). Calculating the orbit of DSG 7Aa,Ab was sufficiently 
challenging that we averaged the astrometry obtained in the two filters before using that as input
for the orbit calculations. This leaves us with only four points to work with in calculating the visual orbital
elements, and therefore the orbit 
obtained here should be viewed as tentative.

In the case of LSC 45 (WDS 07293+1227, HIP 36387), a preliminary orbit is determined using 
the visual orbit code of MacKnight and Horch (2004), which starts with 
a grid search over a user-specified range of each parameter, and then uses a downhill simplex
algorithm to arrive at final parameters that minimize the reduced-$\chi^{2}$ of the least-squares
fit. In the orbit computed here, Gemini points were given twice as high a weight than any points derived
from 4-m class telescopes, as the astrometry at Gemini is more precise. 
Visual representations of all five orbits calculated here are shown in Figure 6.



\subsection{Further Comments on the Individual Systems}

\subsubsection{$\mu$ Cas = WCK 1Aa,Ab = WDS 01083 + 5455}

We update the visual orbit of Drummond \ea (1995) for this system with visual-spectroscopic
orbital elements for this well-known single-lined spectroscopic binary. It is a
somewhat metal-poor system 
([Fe/H] = $-0.68$; Boyajian \ea 2008) with a composite spectral type of
G5V. Both the period and semi-major axis are slightly decreased from the Drummond \ea values,
leading to a lower total mass estimate by a few percent. The {\it Gaia} DR2 does not list a parallax for this system, but
fortunately the
revised {\it Hipparcos} result (van Leeuwen, 2007) is precise to better than 1\%: $132.38 \pm 0.82$ mas. From this we 
determine that $M_{\rm tot} = 0.906 \pm 0.023 M_{\odot}$.

\subsubsection{YSC 129 = WDS 06416+3556}

This system, with period of 2.57 years, was first resolved at the WIYN telescope in 2010 as a part of the Yale-Southern 
Connecticut speckle program. It has a magnitude difference of less than one magnitude
for the wavelengths of observation used here, and is also a known double-lined spectroscopic binary.
While ours
is the first visual orbit based on image data resolving the two stars, there is an 
astrometric orbit in the literature due to Ren and Fu (2010)
where the orbit was determined by examining the intermediate astrometric data from {\it Hipparcos}.
Our orbital parameters essentially confirm and slightly improve their results. The parallax in
the {\it Gaia} DR2 release is given as $\pi = 14.4824 \pm 0.0705$ mas, which implies mass sum
of $1.83 \pm 0.14$ solar masses. The system is slightly metal-poor with an [Fe/H] value of 
$-0.20$. The spectral type given in SIMBAD is F8III; based on the magnitude difference, the secondary is 
probably in the mid-G range. We can also derive a parallax given the spectroscopic and visual
orbital elements; we obtain $13.1 \pm 0.3$ mas, which is almost 10\% lower than the Gaia value.
Given that there are not yet very many Gemini points on this orbit, it is possible that the semi-major
axis could be biased by the 4-m telescope data in the literature. On the other hand, the 
system would not have been resolved by {\it Gaia}, and it is possible that the acceleration of
the system affects the DR2 result. Further large-aperture observations are warranted to 
reduce the uncertainty of the orbital parallax as we wait for future {\it Gaia} updates.
A mass ratio of $0.952 \pm 0.015$ is implied by the orbit; from this and the total mass 
we deduce individual masses of $0.94 \pm 0.07 M_{\odot}$ and $0.89 \pm 0.09 M_{\odot}$.

\subsubsection{LSC 45 = WDS 07293+1227}

The orbit presented here ($P=40.8$ years) is the first visual orbit of this system, which does not have a spectroscopic
orbit in the literature but does appear in the Geneva-Copenhagen catalog with [Fe/H] of $-0.52$ and 
is listed as a double-lined system with mass ratio of $0.935 \pm 0.048$. The system is somewhat
distant, with parallax of $6.5387 \pm 0.0614$ mas provided by DR2. Thus, given our period
and semi-major axis, we deduce a total mass of $1.86 \pm 0.31$ solar masses, and individual
masses of $0.96 \pm 0.17$ and $0.90 \pm 0.19$ $M_{\odot}$. 

\subsubsection{DSG 7Aa,Ab = WDS 16329+0315}

We update our previous orbit that appears in Horch \ea 2015a with a visual-spectroscopic 
orbit; in that work, we only had
two epochs to work with which were only days apart, and thus our orbit, even when
fixing the spectroscopic orbital elements to the values known from the orbit of 
Latham \ea (2002), was somewhat uncertain. To make matters
worse, the system is of extremely small separation throughout the orbital path. Nonetheless, in Table 2, we
have presented two other epochs where the system was measured, and these are consistent
with the $P= 0.62$-year orbit we previously determined. The addition of the new points allows for a better orbit
and a revision of the total mass. Because this system is so challenging, with all measures
below the formal diffraction limit at Gemini, we elected to average the results in the two
DSSI filters from each epoch to hopefully make the orbit more robust. The system has composite 
spectral type of K0V, and is very metal-poor, with
[Fe/H] measured as $-1.4$ by Latham \ea { (1992)}. Our magnitude differences for this system
should be viewed as uncertain for the reasons cited above for small-separation single-lined systems; 
nonetheless, this system will be discussed further Section 4.2.

\subsubsection{DSG 8 = WDS 23485+2539}

The only measures to date of this well-known
double-lined spectroscopic binary with period 3.22 years are those that we have taken
at Gemini.
The improved {\it Gaia} parallax combined with a revised orbit here lead
to a much more precise total mass for the system than in our
previous work; we now find $M_{\rm tot} = 2.21 \pm 0.16 M_{\odot}$. A
mass ratio of $1.01 \pm 0.02$ is implied from our orbit, thus
the individual masses are $1.10 \pm 0.08$ and $1.11 \pm 0.09 M_{\odot}$.
While the spectral type in the literature
is F2IV-V, the system is slightly redder than F2; we treat this 
system as an F5+F5.5 dwarf pair in the next section. The parallax
that we can obtain from the SB+visual orbit presented is in
complete agreement with the {\it Gaia} DR2 value: $14.3 \pm 0.4$ mas
from our orbit versus $14.3551 \pm 0.0532$ mas from {\it Gaia}.

\subsection{An Update on Metal-Poor Stellar Masses}

We have a significantly better understanding of the mass-luminosity relation (MLR) for main sequence stars 
due to recent work of {\it e.g.\ }Torres \ea (2010) and Benedict \ea (2016), but neither of these papers includes
stars that are very low in metallicity. Other studies using long baseline 
optical interferometry have begun to address metallicity (e.g.\ Boyajian 2012a, 2012b, Feiden and 
Chaboyer 2012), but only as low as about [Fe/H]=$-$0.5. Our project at Gemini has been extending 
this significantly on the low-metallicity end, with the specific goal of empirically determining an MLR 
for metal-poor stars for the first time. As the observations in Table 2 make clear, 
we were able to expand the sample to the Southern Hemisphere
beginning in 2016, drawing newly-discovered, small-separation binary systems of interest from the 
active speckle program at the SOAR telescope (see {\it e.g.\ }Tokovinin \ea 2018 and references therein).

In Table 6, we list some further observed properties of the five systems 
identified above. The columns give (1) the discoverer designation;
(2) the {\it Hipparcos} number; (3) the {\it Gaia} parallax from DR2; 
(4) the absolute magnitude obtained from the 
apparent magnitude (which is not listed here) and the parallax, where no reddening
correction was made because these systems are all nearby; (5) the 
(composite) spectral type as it appears in SIMBAD; the metal abundance from 
Holmberg \ea (2009), unless the value
is otherwise marked; (6) the $B-V$ value listed in the {\it Hipparcos}
Catalogue, and (7) the average magnitude difference at 692 nm from 
all available DSSI measures of the target. 

Using the most recent Yale isochrones (Spada \ea 2013), 
we investigated the behavior of stellar mass as a function of
metal abundance at fixed $B-V$ color in our previous work
(Horch \ea 2015a), specifically, Figure 8 of that paper.  While there
is some variation depending on age and the choice of the mixing length parameter,
we found that there was little dependence on a star's color (or equivalently,
effective temperature) over the range of interest here. We argued there that,
because the curves are nearly independent of temperature, it should be 
true that the total mass of a binary star system will follow the same trend
if both stars fall within the spectral range of the simulations.

We are now in a position to update the 2015 results with the orbits from the 
work here.
Most of the orbits used in Horch \ea 2015a are not in need of revision, but 
in Table 7, we have dynamical estimates of the total masses of the systems discussed
in the previous section, which
if we combine with the previous work, gives a total of twelve systems. However, 
DSG 6, which was part of the sample in the 2015 paper,
has been removed here for the reasons stated earlier, leaving us with 11 systems
with which to refine the previous results. 
Besides the new orbital information for these five systems, the recent release of the 
{\it Gaia} DR2 provides much better parallaxes than we previously used in the
majority of cases, and two of the orbits have been subsequently updated
by other authors. 
These new data constrain the dynamical masses much better than
was possible with the earlier results. We can therefore combine
the two samples, update the parallaxes and total masses, and arrive at
an improved relationship between total mass and metallicity.

The sample that we can report on now covers a range from the solar value to as low as
[Fe/H]= $-1.4$. We are studying systems that are as metal poor as [Fe/H]=$-2.01$,
but some of these do not all have orbital information of sufficient quality to include in this analysis.
Given the measured properties of each of our binaries, 
the stellar models allow us to estimate
the total mass of the solar-abundance binary star of the same color. We can
compute the ratio of the dynamical masses we determine to the solar analogue for 
the system and examine whether the trend is similar to what the
stellar models predict. To make this mass estimate, we use the composite $B-V$
color for the system as it appears in the {\it Hipparcos} Catalogue
and the average magnitude difference measured by DSSI at 692 nm, 
combining all observations of the system in that filter to date at Gemini, 
{\it i.e.\ }the last two columns of Table 6. As in the previous work, 
we then use the solar-abundance spectral library of Pickles (1998) to 
combine stars of different spectral types to produce a composite $B-V$ value
and $\Delta m$ at 692 nm that is as close as possible to the measured values.
We incorporate into these models a standard atmospheric transmission curve,
the known filter transmission curve, and dichroic transmission curve
for DSSI. 
Table 7 repeats the identification of the targets in the first two columns, and then shows 
in the third and fourth columns the
assigned component spectral types and the composite (system) $B-V$ that were obtained in
this way. If the modeling is reasonably accurate, then the latter should be in good agreement
with the measured $B-V$ for the system
found in the literature and reported in Table 6. This is true in all cases.

Finally, the $B-V$ value associated with the assigned spectral type
of each component is converted into a mass estimate
using the standard reference of Schmidt-Kaler (1982), and
these appear in Columns 6 and 7 of Table 7. 
We have estimated the uncertainty of
the total masses shown by using the scatter in the 
$B-V$ color. 
We also checked the conversion from the photometric information to mass using the information
provided in the work of Boyajian \ea (2012a, 2012b), and found 
very good agreement with the Schmidt-Kaler
reference over the spectral range of interest, and assuming Solar metallicity for our
models. Finally, in the last column of Table 7, we show the dynamical total mass
estimate using Kepler's harmonic law and the data in Table 5. While it is
not used in the analysis here, it is worth noting that for the systems that are 
double-lined spectroscopic binaries, the mass ratios $m_{2}/m_{1}$ implied from
Table 7 are in reasonably good agreement from the values implied from the
spectroscopic orbits in both cases.

In Figure 7, we plot both the theoretical and observed ratio of binary system mass
to the solar-abundance value, as a function of
metal abundance.
We have assumed an uncertainty in metallicity of the observed data of 0.1 dex
(which is the uncertainty stated in Holmberg \ea [2009], 
the source of most of our abundance values).
The plot suggests that, within the uncertainty, the points follow the 
trend expected from the stellar structure calculations. In this updated
plot, we have been able to shrink the vertical error 
bars by roughly 25 to 50\% in general and therefore make a more definitive statement than before concerning the
good agreement between the observational data and stellar models for a wide range of
metallicity, given the observational uncertainties. Furthermore, it is interesting to note that the data points usually lie closest to the 
model curves for the correct spectral type.
While Figure 7 falls short of a true metal-poor MLR, it does allow us to include systems where no mass ratio information is 
available or when it is of low quality (as it is for most of the single-lined spectroscopic binaries in the sample). 
Thus, the value of the plot is as a tool to investigate the behavior of mass as a function of metallicity
with the highest precision information that we have at present, namely the total mass of each system. When we 
have determined individual masses of high precision for more of our sample and over a greater range of metallicity, we hope to 
construct an empirical metal-poor MLR that would give a relevant comparison to theoretical stellar models.

\subsection{Triple Stars}

Of the twelve new components listed in Table 2, seven were stars observed on the 
metal-poor program, two were part of the triple stars program, and three were 
serendipitous discoveries where the primary star was originally used as a point source 
calibration object. In both
cases for the triple star program, the architecture of the system is Aa,Ab-B. For the 
metal poor program, of the seven discoveries, two have single-lined orbits in
the literature (HIP 4754 and HIP 80925), one has a double-lined orbit (HIP 81170),
and four have mass ratios listed in the Geneva-Copenhagen spectroscopic catalog
(HIP 19915, HIP 56851, HIP 57421, and HIP 80925), indicating that a double-lined
orbit would be possible. We can infer from this
that four of the seven systems are in fact triple stars, because in three cases
the separations we measure
are too large to correspond to the spectroscopic component and in the fourth, HIP 72492,
we see explicitly that the system is triple in Figure 2. HIP 72492 is the sole example
in this set of discoveries that is of architecture A-Ba,Bb. The seven systems span
a very large range in metallicity; HIP 57421 has [Fe/H] of $+0.15$ and HIP 19915
has [Fe/H] of $-2.01$ as shown in the Geneva-Copenhagen catalog.

Considering all of the triple stars in Table 2, the  observations  of  
hierarchical  multiple systems  presented  here
provide  rich material for  the study  of their  orbital architecture.
Triple systems where  both outer and inner orbits  are known and their
relative   orientation  can   be  determined   are   still  relatively
rare.  The calculation  of  such  orbits  is outside  the  scope  of  this
paper, but to illustrate the  power of DSSI at Gemini, we provide
in Figure 8 the  inner orbit of the 5-year subsystem in WDS 16057$-$0617
(HIP  78849). We  comment  below on  some  remarkable hierarchies,  in
addition to the comments on new resolutions provided in Section~3.1.

\noindent
$\bullet$ WDS 11221$-$2447  (HIP  55505)  is  the  young  pre-main  sequence  quadruple
TWA~4. The double-lined subsystem Ba,Bb is resolved here. It was previously
resolved only with the Keck interferometer. 

\noindent
$\bullet$ WDS 11596$-$7813 (HIP 58484) is the young triple system in the
$\epsilon$~Cha association composed of three B9V stars. Its eccentric inner
orbit will soon be determined (see Figure 5 in Brice\~{n}o and Tokovinin, 2017).

\noindent
$\bullet$ WDS 14598$-$2201 (HIP 73385). The astrometric subsystem with a period of
3.55 years is not resolved here. Its previous resolutions at SOAR (TOK~47 Aa,Ab) might
be spurious.  This is a metal-poor triple system. 

\noindent
$\bullet$ WDS 16120$-$1928  (HIP  79374).  This  is  the  young hierarchical  system
$\nu$~Sco  with 7 components  in the  Upper Scorpius  association. The
resolved tight pair CHR~146 Aa,Ab has an estimated period of less than
10 years but no orbit, so far, owing to the lack of speckle coverage. 

\noindent
$\bullet$ WDS 16253$-$4909 (HIP 80448) is a young hierarchical quadruple system. We
resolved the subsystem CVN~27 Ba,Bb, for which a
20-yr orbit has been computed (Tokovinin, 2018). 
Our measurement is in excellent agreement with this
orbit. 

\noindent
$\bullet$ WDS 17157$-$0949 (HIP  84430) is an  interesting triple system  where both
inner and outer  orbits are known, with periods of  5.1 and 150 years,
respectively.  The orbits  are  not coplanar.  The dynamical  parallax
deduced from  the orbits, about  7 mas, disagrees with  both {\it Hipparcos}
(5.7 mas) and {\it Gaia} (2.7 mas).  

\noindent
$\bullet$ WDS 18112$-$1951 (HIP  89114) has the  inner subsystem TOK~57  Aa,Ab (also
known from  lunar occultations) with  a short ($\sim$10  years) period
but still undefined orbit, to which our measurement contributes.

\noindent
$\bullet$ WDS 18126$-$7430  (HIP   89234)  is   a  triple  system   with  comparable
separations  (potentially  interacting)  and  a known  324-year  outer
orbit. The measurement made here  allows us to compute the preliminary
inner  orbit of  TOK~58  Aa,Ab with  a  period of  39  years and  high
eccentricity. More  data are needed to  confirm this orbit  and make a
full dynamical study of this system.

\noindent
$\bullet$ WDS 19453$-$6823 (HIP 97196)  is similar to the previous  object, having
comparable separations  of the inner and outer  subsystems.  The outer
orbit  is   not  known  (estimated  period  40   years),  while  these
observations  allow  us to  compute  the  preliminary  inner orbit  of
TOK~425 Ba,Bb with a period of 4 years.

\section{Conclusions}

We have presented { 248} 
measures of close binary and trinary stars where the data 
were taken at the Gemini North and South Telescopes over the last three years
using speckle imaging. The astrometric and photometric precision of this sample
appears similar to previous papers, namely $\sim$1.3 mas in separation and
$\sim$0.6$^{\circ}$ in position angle { when measures in both instrument channels are averaged}, 
and $\sim$0.1 mag in magnitude difference
in general. However, these numbers increase somewhat when the separation
is below the diffraction limit and/or the magnitude difference is above 4. { The overall
accuracy of our measures is affected by the uncertainty in the scale and orientation values;
we find that the pixel scales used here are uncertain at approximately the 0.4\% level, and that the
position angle offsets are uncertain to 0.4$^{\circ}$. To judge the overall
uncertainty of any given measure, these numbers should be added in quadrature with
the internal precision numbers. However, no systematic offsets in position angle or
separation were noted in studying systems where independent positional information
was available.}
Twelve previously unknown companions were discovered.

Given these measures, we computed visual orbital elements for 5 systems, one of which had
no previous visual orbit (although it was known to be a spectroscopic binary), and
the other four of which were calculated as a visual-spectroscopic orbit.
Using these orbits, our previous work, and recent parallax results from {\it Gaia}, we found it possible 
to refine our understanding of the effect of metallicity on the total mass of a binary, and
conclude that the trend in lower mass for a given spectral type for metal-poor systems
predicted by stellar models remains borne out in these observations. Future observations
of this kind could allow us to construct an empirical low-metallicity mass-luminosity relation
and also to understand the role of metallicity in stellar multiplicity statistics.

Including the discovered components, we resolved 19 systems into triples and a number of
other objects we report on here are sub-systems of known triples or multiples. These results
have allowed us to continue the work of characterizing these systems as a part of a
broader census of triples that could tell us more about star formation mechanisms in the
future.

\acknowledgments
We gratefully acknowledge support from the National Science Foundation, Grants
AST 15-17824 and AST 16-16698, in the completion of this work. We also thank the
excellent staff at both Gemini telescopes for their help and support during our observing
runs. We are mindful of the fact that the Mauna Kea is a sacred site for native Hawaiians
and are grateful for the opportunity to be present there.
E.H. thanks W. van Altena and P. Demarque for their continued interest in
and support of this work. { We thank the anonymous referee for providing
a number of comments and corrections that improved the paper.}
These observations were taken as a part of NOAO programs
2015B-0081, 2016A-0225, and 2017A-0205. We made use of
the Washington Double Star Catalog maintained at the U.S. Naval Observatory, the
SIMBAD database, operated at CDS, Strasbourg, France, the Ninth Catalogue of 
Spectroscopic Orbits of Binary Stars, maintained by D. Pourbaix. 
It also made use of data from the European Space Agency (ESA) mission
{\it Gaia} (\url{https://www.cosmos.esa.int/gaia}), processed by the {\it Gaia}
Data Processing and Analysis Consortium (DPAC,
\url{https://www.cosmos.esa.int/web/gaia/dpac/consortium}). Funding for the DPAC
has been provided by national institutions, in particular the institutions
participating in the {\it Gaia} Multilateral Agreement.




\begin{centering}

\begin{deluxetable}{lccclrl}
\tabletypesize{\scriptsize}
\tablewidth{0pt}
\tablenum{1}
\tablecaption{Orbits Used in the Scale Determinations}
\tablehead{
\colhead{Run} &
\colhead{Telescope} &
\colhead{Instrument} &
\colhead{WDS} &
\colhead{Discoverer} &
\colhead{HIP} &
\colhead{Orbit Reference} \\
&&&& \colhead{Designation} 
}

\startdata

2015 July & Gemini-N & DSSI & $17080+3556$ & HU 1176AB & 83838 & Muterspaugh \ea (2010)\\
 &&& $21145+1000$ & STT 535AB & 104858 & Muterspaugh \ea (2008)\\

2016 January & Gemini-N & DSSI  & $04136+0743$ & A  1938 & 19719 & Muterspaugh \ea (2010)\\
 &&& $22409+1433$ &  HO  296AB & 111974 & Muterspaugh \ea (2010)\\

2016 June  & Gemini-S & DSSI & $18384-0312$ & A 88AB & 91394 &  Griffin (2013)\\
  &&& $19026-2953$ & HDO 150AB & 93506 & DeRosa \ea (2012)\\

2017 June & Gemini-S & DSSI & $18384-0312$ & A 88AB & 91394 & Griffin (2013)\\
  &&& $19026-2953$ & HDO 150AB & 93506 & DeRosa \ea (2012)\\
 
 2017 December & Gemini-N & 'Alopeke & $13100+1732$ & STF 1728AB & 64241 & Muterspaugh \ea (2015) \\
  &&& $15232+3017$ & STF 1937AB & 75312 & Muterspaugh \ea (2010) \\
  &&& $15278+2906$ & JEF 1 & 75695 & Muterspaugh \ea (2010)\\
 
 2018 March/April & Gemini-S & DSSI & $06573-3530$ & I 65 & 33451 & Docobo \& Ling (2009) \\
  &&& $07518-1354$ & BU  101 & 38382 & Tokovinin (2012) \\
  &&& $08270-5242$ & B 1606 & 41426 & Tokovinin \ea (2015) \\
  &&& $16044-1122$ & STF 1998AB & 78727 & Docobo \& Ling (2009) \\
  
 2018 March/April & Gemini-N & 'Alopeke & $13100+1732$ & STF 1728AB & 64241 & Muterspaugh \ea (2015) \\
  &&& $15232+3017$ & STF 1937AB & 75312 & Muterspaugh \ea (2010) \\
  &&& $15278+2906$ & JEF 1 & 75695 & Muterspaugh \ea (2010) \\
 
 \enddata
 
 \end{deluxetable}     

\pagestyle{empty}
\begin{deluxetable}{lllrlrrrrl}
\tabletypesize{\scriptsize}
\tablenum{2}
\tablecaption{Binary star speckle measures}
\tablehead{ 
\colhead{WDS} & 
\colhead{HR,ADS} & 
\colhead{Discoverer} & HIP & 
\colhead{Date} & 
 \colhead{$\theta$} & \colhead{$\rho$} & 
\colhead{$\Delta m$} & 
\colhead{$\lambda$} & 
\colhead{$\Delta \lambda$} \\
\colhead{($\alpha$,$\delta$ J2000.0)} & 
\colhead{DM,etc.} & \colhead{Designation} && 
\colhead{(2000+)} & 
\colhead{($^{\circ}$)} & 
\colhead{(${\prime \prime}$)} 
 & \colhead{(mag)}
 & \colhead{(nm)} & 
\colhead{(nm)} 
} 
\startdata 

$01011+1622$ & BD$+$15 150 & DSG 9 & 4754 & 
   16.0338 &  36.0 & 1.3607 & $<$5.11\tablenotemark{a}  & 880 & 50 \\
&&&&
   16.0420 &  34.0 & 1.3479 &  $<$6.24\tablenotemark{a} & 692 & 40 \\
&&&&
   16.0420 &  35.8 & 1.3484 &  $<$7.74\tablenotemark{a} & 880 & 50 \\
&&&&
   16.0448 &  35.0 & 1.3678 &  $<$5.94\tablenotemark{a} & 880 & 50 \\
&&&&
   17.9476 &  38.0 & 1.3679 &  $<$6.48\tablenotemark{a} & 832 & 40 \\

$01057+2128$ & ADS 899 & YR 6Aa,Ab &    5131 &    
   17.9476 & 311.8 & 0.1404 & 0.87 & 562 & 44 \\
&&&&
   17.9476 & 312.0 & 0.1405 & 0.67 & 832 & 40 \\

$01083+5455$ & HD 6582 & WCK 1Aa,Ab &    5336 &   
   15.5448 &  34.5 & 0.9018 & 5.13 & 692 & 40 \\
&&&&
   15.5448 &  32.0 & 0.8991 & 4.69 & 880 & 50 \\
&&&&
   16.0337 &  30.9 & 0.7560 & 7.16 & 692 & 40 \\
&&&&
   16.0337 &  31.0 & 0.7510 & 5.51 & 880 & 50 \\
&&&&
   16.0474 &  30.6 & 0.7513 & 4.81 & 692 & 40 \\
&&&&
   16.0474 &  27.7 & 0.7589 & 4.42 & 880 & 50 \\

$02396-1152$ & HR 781 & FIN 312 &   12390 &       
   17.9477 & 203.1 & 0.0830 & 0.71 & 562 & 44 \\
&&&&
   17.9477 & 203.6 & 0.0831 & 0.72 & 832 & 40 \\

$02422+4012$ & HR 788 & MCA 8 &   12623 &  
   16.0395 & 61.2 & 0.0245 & 0.33 & 692 & 40 \\
&&&&
   16.0395 & 55.4 & 0.0273 & 0.42 & 880 & 50 \\
&&&&
   16.0475 &  49.1 & 0.0320 & 0.61 & 692 & 40 \\
&&&&
   16.0475 &  49.0 & 0.0310 & 0.48 & 880 & 50 \\

$04107-0452$ & ADS 3041 & A 2801 &   19508 &  
   16.0368 & 346.3 & 0.1274 & 0.73 & 692 & 40 \\
&&&&
   16.0368 & 346.1 & 0.1276 & 0.65 & 880 & 50 \\
&&&&
   16.0476 & 346.4 & 0.1281 & 0.76 & 692 & 40 \\
&&&&
   16.0476 & 346.3 & 0.1285 & 0.77 & 880 & 50 \\
&&&&
   18.2460 & 355.9 & 0.1641 & 0.72 & 562 & 44 \\
&&&&
   18.2460 & 355.4 & 0.1618 & 0.85 & 832 & 40 \\

$04136+0743$ & ADS 3064 & A 1938 &   19719 &    
   17.9477 & 329.8 & 0.1399 & 0.88 & 562 & 44 \\
&&&&
   17.9477 & 330.3 & 0.1395 & 0.74 & 832 & 40 \\

$04163+3644$ & HD  26872 & YSC 128AB &   19915 & 
   16.0422 & 124.2 & 0.0133 & 1.07 & 692 & 40 \\
&&&&
   16.0422 & 120.1 & 0.0071 & 0.41 & 880 & 50 \\
&&&&
   16.0476 & 147.8 & 0.0113 & 1.19 & 692 & 40 \\
&&&&
   16.0476 & 110.3 & 0.0067 & 0.17 & 880 & 50 \\
$04163+3644$ & HD  26872 & DSG 10AC &   19915 & 
   16.0422 & 224.7 & 0.5286 & 7.37 & 692 & 40 \\
&&&&
   16.0422 & 219.0 & 0.5352 & 4.81 & 880 & 50 \\
&&&&
   16.0476 & 220.1 & 0.5255 & 5.56 & 692 & 40 \\
&&&&
   16.0476 & 215.2 & 0.5247 & 5.39 & 880 & 50 \\

$06416+3556$ & HD  47703 & YSC 129 &   32040 & 
   16.0371 & 274.7 & 0.0299 & 0.73 & 692 & 40 \\
&&&&
   16.0371 & 275.5 & 0.0303 & 0.78 & 880 & 50 \\
&&&&
   16.0477 & 271.9 & 0.0279 & 0.72 & 692 & 40 \\
&&&&
   16.0477 & 275.4 & 0.0301 & 0.84 & 880 & 50 \\

$07043-5645$ & HR 2683 & DSG 11 & 34105 &
   17.4321 & 296.9 & 0.0645 & 1.09 & 692 & 40 \\
&&&&
   17.4321 & 296.5 & 0.0660 & 1.00 & 880 & 50 \\

$07293+1227$ & HD  59179 & LSC 45 &   36387 & 
   16.0398 & 246.7 & 0.0588 & 1.43 & 692 & 40 \\
&&&&
   16.0398 & 246.7 & 0.0592 & 1.43 & 880 & 50 \\
&&&&
   16.0480 & 246.9 & 0.0603 & 1.44 & 692 & 40 \\
&&&&
   16.0480 & 246.6 & 0.0584 & 1.55 & 880 & 50 \\
&&&&
   18.2515 & 279.4 & 0.0460 & 1.45 & 562 & 44 \\
&&&&
   18.2515 & 279.9 & 0.0495 & 1.56 & 832 & 40 \\

$07546-0125$ & HD  64606  & YSC 198Aa,Ab &   38625 & 
   16.0399 & 328.2 & 0.7711 & 3.96 & 692 & 40 \\
&&&&
   16.0399 & 328.1 & 0.7721 & 3.11 & 880 & 50 \\
&&&&
   16.0480 & 328.2 & 0.7715 & 3.98 & 692 & 40 \\
&&&&
   16.0480 & 328.1 & 0.7723 & 3.08 & 880 & 50 \\

$11221-2447$ & HD  98800 & I   507A,Ba &   55505 &    
   17.4295 &   4.9 & 0.4914 & 0.15 & 692 & 40 \\
&&&&
   17.4295 &   4.8 & 0.4917 & 0.00 & 880 & 50 \\
$11221-2447$ & HD  98800 & BOD   1Ba,Bb &   55505 &  
   17.4295 & 288.8 & 0.0172 & 0.17 & 692 & 40 \\
&&&&
   17.4295 & 274.0 & 0.0174 & 0.05 & 880 & 50 \\

$11394-3923$ & HD 101309  & DSG 12 &   56851 & 
   18.2429 & 103.4 & 0.0315 & 2.35 & 692 & 40 \\ 
&&&&
   18.2429 & 109.4 & 0.0315 & 2.93 & 880 & 50 \\

$11464-2758$ & HD 102301 & LSC 49AB &   57421 & 
   16.0348 &  80.2 & 0.2097 & 2.84 & 692 & 40 \\
&&&&
   16.0348 &  80.5 & 0.2104 & 2.80 & 880 & 50 \\
&&&&
   16.0429 &  80.3 & 0.2101 & 2.72 & 692 & 40 \\
&&&&
   16.0429 &  80.4 & 0.2100 & 2.68 & 880 & 50 \\
&&&&
   17.4293 &  73.5 & 0.2055 & 2.80 & 692 & 40 \\
&&&&
   17.4293 &  73.4 & 0.2057 & 2.66 & 880 & 50 \\
&&&&
   18.2430 &  69.8 & 0.2010 & 2.78 & 692 & 40 \\
&&&&
   18.2430 &  69.5 & 0.2009 & 2.65 & 880 & 50 \\
&&&&
   18.2438 &  69.7 & 0.2040 & 2.90 & 562 & 44 \\
&&&&
   18.2438 &  70.7 & 0.2005 & 2.66 & 832 & 40 \\
$11464-2758$ & HD 102301 & DSG 13AC &   57421 & 
   16.0429 & 152.6 & 0.4764 & 4.20 & 692 & 40 \\
&&&&
   16.0429 & 154.1 & 0.4764 & 4.77 & 880 & 50 \\
&&&&
   18.2430 & 150.8 & 0.4638 & 7.06 & 692 & 40 \\
&&&&
   18.2430 & 151.7 & 0.4683 & 4.88 & 880 & 50 \\

$11596-7813$ & HD 104174 & HJ 4486Aa,Ab &   58484 &  
   17.4296 &  33.1 & 0.0549 & 0.19 & 692 & 40 \\
&&&&
   17.4296 &  32.5 & 0.0568 & 0.19 & 880 & 50 \\
$11596-7813$ & HD 104474 & HJ 4486Aa,B &   58484 & 
   17.4296 & 245.8 & 0.1712 & 0.42 & 692 & 40 \\
&&&&
   17.4296 & 245.8 & 0.1707 & 0.15 & 880 & 50 \\

$12018-3439$ & HD 104471 & DSG 14Aa,Ab &   58669 & 
   17.4323 & 259.0 & 0.0152 & 0.02 & 692 & 40 \\
&&&&
   17.4323 & 255.1 & 0.0153 & 0.02 & 880 & 50 \\
$12018-3439$ & HD 104471 & I 215Aa,B &   58669 & 
   17.4323 & 345.0 & 0.1895 & 0.60 & 692 & 40 \\
&&&&
   17.4323 & 345.0 & 0.1895 & 0.47 & 880 & 50 \\

$12319-6330$ & HR 4771 & DSG 15 & 61158 &
   17.4324 &  48.9 & 0.0831 & 3.10 & 692 & 40 \\
&&&&
   17.4324 &  49.6 & 0.0814 & 3.08 & 880 & 50 \\
   
$12565-2635$ & HD 112375  & YSC 216Aa,Ab &   63162 & 
   16.0404 &  18.2 & 0.1529 & 0.85 & 692 & 40 \\
&&&&
   16.0404 &  18.1 & 0.1530 & 1.16 & 880 & 50 \\
&&&&
   16.0459 &  18.3 & 0.1528 & 0.75 & 692 & 40 \\
&&&&
   16.0459 &  18.2 & 0.1526 & 1.00 & 880 & 50 \\
&&&&
   17.4320 &  25.8 & 0.1463 & 1.05 & 692 & 40 \\
&&&&
   17.4320 &  25.7 & 0.1468 & 1.07 & 880 & 50 \\
&&&&
   18.2429 &  31.5 & 0.1425 & 0.96 & 692 & 40 \\
&&&&
   18.2429 &  31.4 & 0.1429 & 1.01 & 880 & 40 \\

$13495-2621$ & HD 120368  & TOK 405 &   67458 & 
   16.4906 & 193.7\tablenotemark{b} & 0.7168 & 5.23 & 562 & 40 \\
&&&&
   16.4906 & 194.2\tablenotemark{b} & 0.7234 & 4.57 & 880 & 50 \\
&&&&
   17.4323 & 192.8\tablenotemark{b} & 0.7080 &  $<$5.47\tablenotemark{a} & 692 & 40 \\
&&&&
   17.4323 & 192.7\tablenotemark{b} & 0.7105 &  $<$4.28\tablenotemark{a} & 880 & 50 \\
&&&&
   18.2456 & 192.0\tablenotemark{b} & 0.6953 & 5.03 & 692 & 40 \\
&&&&
   18.2456 & 192.6\tablenotemark{b} & 0.7006 & 4.34 & 880 & 50 \\

$14025-2440$ & HD 122445 & DSG 16Aa,Ab &   68587 &  
   17.4297 & 296.8 & 0.0089 & 0.08 & 692 & 40 \\
&&&&
   17.4297 & 294.3 & 0.0103 & 0.06 & 880 & 50 \\
$14025-2440$ & HD 122445 & B 263Aa,B &   68587 & 
   17.4297 & 126.8 & 0.1983 & 1.33 & 692 & 40 \\
&&&&
   17.4297 & 126.9 & 0.1974 & 1.28 & 880 & 50 \\

$14494-5726$ &HD 130264  & HDS 2092AB &   72492 &   
   17.4325 & 258.3 & 0.3754 & 3.62 & 692 & 40 \\
&&&&
   17.4325 & 258.2 & 0.3759 & 3.53 & 880 & 50 \\
$14494-5726$ & HD 130264 & HDS 2092BC &   72492 &   
   17.4325 & 186.4 & 0.0621 & 0.46 & 692 & 40 \\
&&&&
   17.4325 & 186.1 & 0.0621 & 0.30 & 880 & 50 \\

$14509-1603$ & HD 130841 & DSG 17Aa,Ab &   72622 &  
   17.4438 & 347.9 & 0.0283 & 0.65 & 692 & 40 \\
&&&&
   17.4438 & 348.4 & 0.0275 & 0.37 & 880 & 50 \\
&&&&
   17.4462 & 350.7 & 0.0268 & 0.28 & 692 & 40 \\
&&&&
   17.4462 & 350.7 & 0.0272 & 0.39 & 880 & 50 \\

$14598-2201$ &HD 132475  & TOK 47AB &   73385 &  
   17.4325 & 130.2 & 1.0875 &  $<$5.86\tablenotemark{a} & 692 & 40 \\
&&&&
   17.4325 & 130.4 & 1.0862 &  $<$5.46\tablenotemark{a} & 880 & 50 \\

$15006+0836$ & HD 132756 & YSC 8 &   73449 &  
   17.4407 & 150.7 & 0.0805 & 0.23 & 692 & 40 \\
&&&&
   17.4407 & 150.8 & 0.0806 & -0.16\tablenotemark{d} & 880 & 50 \\

$15317+0053$ & HD 138369 & TOK 48 &   76031 &   
   17.4408 & 106.6 & 0.0517 & 1.46 & 692 & 40 \\
&&&&
   17.4408 & 104.2 & 0.0503 & 1.23 & 880 & 50 \\

$15362-0623$ & HD 139059 & TOK 301Aa,Ab &   76400 & 
   16.4906 &  37.0 & 0.1743 & 5.74 & 562 & 40 \\
&&&&
   16.4906 &  40.4 & 0.1803 & 4.15 & 880 & 50 \\
&&&&
   17.4408 &  25.5 & 0.1831 & 5.14 & 692 & 40 \\
&&&&
   17.4408 &  24.6 & 0.1856 & 4.00 & 880 & 50 \\
&&&&
   18.2490 &  12.8 & 0.1938 & 4.66 & 692 & 40 \\
&&&&
   18.2490 &  11.1 & 0.1986 & 3.22 & 880 & 50 \\

$15365+1607$ & HR 5804 & DSG 18Aa,Ab & 76424 &
    14.5637 & 320.4 & 1.0195 & $<$7.00\tablenotemark{a} & 692 & 40 \\
&&&&
    14.5637 & 320.1 & 1.0215 & 5.27 & 880 & 50 \\
&&&&
    15.5248 & 319.9 & 1.0095 & $<$6.49\tablenotemark{a} & 692 & 40 \\
&&&&
    15.5248 & 320.1 & 1.0098 & 5.53 & 880 & 50 \\
&&&&
    15.5300 & 320.7 & 1.0036 & 6.08 & 692 & 40 \\
&&&&
    15.5300 & 320.1 & 1.0082 & 5.92 & 880 & 50 \\
&&&&
    15.5330 & 320.3 & 1.0061 & $<$6.27\tablenotemark{a} & 692 & 40 \\
&&&&
    15.5330 & 320.1 & 1.0144 & $<$5.49\tablenotemark{a} & 880 & 50 \\
&&&&
    15.5409 & 321.3 & 1.0071 & $<$7.35\tablenotemark{a} & 692 & 40 \\
&&&&
    15.5409 & 320.1 & 1.0079 & $<$5.59\tablenotemark{a} & 880 & 50 \\
&&&&
    15.5437 & 320.7 & 1.0162 & 6.74 & 692 & 40 \\
&&&&
    15.5437 & 320.1 & 1.0065 & 5.06 & 880 & 50 \\

$16035-5747$ & HR 5961 & SEE 258AB &   78662 &     
   16.4906 & 320.1\tablenotemark{b} & 0.1896 & 0.32 & 562 & 40 \\
&&&&
   16.4906 & 320.3\tablenotemark{b} & 0.1892 & 0.23 & 880 & 50 \\
&&&&
   17.4409 & 102.7 & 0.1626 & 0.27 & 692 & 40 \\
&&&&
   17.4409 & 102.6 & 0.1622 & 0.10 & 880 & 50 \\
&&&&
   18.2490 & 245.6\tablenotemark{b} & 0.1568 & 0.00 & 692 & 40 \\
&&&&
   18.2490 & 245.4\tablenotemark{b} & 0.1568 & 0.21 & 880 & 50 \\

$16044-1122$ & ADS 9909  & STF 1998AB &   78727 &   
   18.2490 & 189.1\tablenotemark{b} & 1.1216 & 0.43 & 692 & 40 \\
&&&&
   18.2490 & 188.9\tablenotemark{b} & 1.1238 & 0.45 & 880 & 50 \\

$16057-0617$ & HD 144362 & FIN 384Aa,Ab &   78849 &  
   17.4381 &  20.1\tablenotemark{b} & 0.0277 & 0.02 & 692 & 40 \\
&&&&
   17.4381 &  20.9\tablenotemark{b} & 0.0275 & 0.02 & 880 & 50 \\
$16057-0617$ & HD 144362 & BU  948Aa,B &   78849 & 
   17.4381 &  77.8 & 0.6866 & 1.77 & 692 & 40 \\
&&&&
   17.4381 &  77.7 & 0.6868 & 1.53 & 880 & 50 \\

$16120-1928$ & ADS  9951 & CHR 146Aa,Ab &   79374 &  
   17.4298 & 346.1 & 0.0343 & 2.15 & 692 & 40 \\
&&&&
   17.4298 & 346.1 & 0.0359 & 1.92 & 880 & 50 \\
$16120-1928$ & ADS  9951 & BU  120Aa,B &   79374 & 
   17.4298 &   1.0 & 1.3317 &  $<$1.34\tablenotemark{a} & 692 & 40 \\
&&&&
   17.4298 &   1.0 & 1.3368 &  $<$1.20\tablenotemark{a} & 880 & 50 \\

$16142-5047$ & HD 145598 & TOK 409 &   79576 & 
   17.4409 &  26.1\tablenotemark{b?} & 0.0171 & 1.26 & 692 & 40 \\
&&&&
   17.4409 &  17.8\tablenotemark{b?} & 0.0115 & 1.24 & 880 & 50 \\
&&&&
   18.2490 & 352.0 & 0.0189 & 2.28 & 692 & 40 \\
&&&&
   18.2490 & 158.6\tablenotemark{b?} & 0.0132 & 2.24 & 880 & 50 \\

$16253-4909$ & HD 147633B & CVN  27Ba,Bb &   80448B &  
   17.4299 & 204.4 & 0.0675 & 2.99 & 692 & 40 \\
&&&&
   17.4299 & 204.8 & 0.0684 & 2.91 & 880 & 50 \\

$16315-3901$ & HD 148704 & DSG 19 &   80925 & 
   16.4906 & 168.0 & 0.0089 & 0.64 & 562 & 40 \\
&&&&
   16.4906 & 172.1 & 0.0082 & 0.51 & 880 & 50 \\
&&&&
   17.4409 & 179.2 & 0.0117 & 0.69 & 692 & 40 \\
&&&&
   17.4409 & 178.9 & 0.0137 & 0.63 & 880 & 50 \\

$16329+0315$ & HD 149162 & DSG 7 Aa,Ab & 81023 &    
   15.5276 & 236.6 & 0.0131 & ...\tablenotemark{c} & 692 & 40 \\ 
&&&&
   15.5276 & 256.1 & 0.0097 & ...\tablenotemark{c}  & 880 & 50 \\ 
&&&&
   18.2468 & 354.0 & 0.0149 & ...\tablenotemark{c}  & 562 & 44 \\ 
&&&&
   18.2468 &  15.5 & 0.0072 & ...\tablenotemark{c}  & 832 & 40 \\ 
$16329+0315$ & HD 149162 & DSG 7 Aa,Ac & 81023 &
   15.5276 & 233.3 & 0.2970 & 4.93 & 692 & 40 \\
&&&&
   15.5276 & 233.2 & 0.2934 & 3.90 & 880 & 50 \\
&&&&
   18.2468 & ... & ... & $>$5.50\tablenotemark{e} & 562 & 44 \\
&&&&
   18.2468 & 242.6 & 0.2677 & 3.95 & 832 & 40 \\

$16347-0414$ & HD 149414 & DSG 20 &   81170 & 
   15.5412 & 205.8 & 0.0147 & 1.13 & 692 & 40 \\
&&&&
   15.5412 & 206.7 & 0.0169 & 2.29 & 880 & 50 \\

$17066+0039$ & ADS 10341 & BU 823A,Ba &   83716 & 
   17.4327 & 169.8 & 1.0088 &  $<$2.02\tablenotemark{a} & 692 & 40 \\
&&&&
   17.4327 & 169.8 & 1.0126 &  $<$1.68\tablenotemark{a} & 880 & 50 \\
$17066+0039$ & ADS 10341 & TOK 52Ba,Bb &   83716 & 
   17.4327 & 228.3 & 0.0775 & 0.01 & 692 & 40 \\
&&&&
   17.4327 & 228.7 & 0.0796 & 0.00 & 880 & 50 \\

$17157-0949$ & ADS 10423 & A 2592A,Ba &   84430 & 
   17.4410 &  91.2 & 0.1862 & 1.30 & 692 & 40 \\
&&&&
   17.4410 &  91.1 & 0.1861 & 1.66 & 880 & 50 \\
$17157-0949$ & ADS 10423 & TOK 53Ba,Bb &   84430 & 
   17.4410 & 221.3 & 0.0350 & 0.32 & 692 & 40 \\
&&&&
   17.4410 & 220.9 & 0.0352 & 0.30 & 880 & 50 \\

$17247+3802$ & BD+38 2932 & HSL   1Aa,Ab & 85209 &  
  18.2496 &  55.8 & 0.0232 & 0.23 & 562 & 44 \\
&&&&
  18.2496 &  52.2 & 0.0243 & 0.42 & 832 & 40 \\

$17247+3802$ & BD+38 2932 & HSL   1Aa,Ac & 85209 &  
   15.5248 &  58.6 & 0.2486 & 2.70 & 692 & 40 \\
&&&&
   15.5248 &  58.8 & 0.2464 & 2.45 & 880 & 50 \\
&&&&
   18.2496 &  56.8 & 0.3176 & 2.43 & 562 & 44 \\
&&&&
      18.2496 &  56.7 & 0.3138 & 1.88 & 832 & 40 \\

$17341-0303$ & HD 159307 & TOK 417 &   85963 & 
   18.2496 &   2.3 & 0.0444 & 2.95 & 832 & 40 \\

$17362-1752$ & HD 159589 & YSC 158Aa,Ab  &   86142 &  
   17.4437 &   2.1 & 0.0153 & 0.33 & 692 & 40 \\
&&&&
   17.4437 &   3.8 & 0.0172 & 0.54 & 880 & 50 \\
$17362-1752$ & HD 159589 & HDS2485Aa,B  &   86142 & 
   17.4437 &  91.0 & 0.7413 & 2.41 & 692 & 40 \\
&&&&
   17.4437 &  91.1 & 0.7406 & 2.54 & 880 & 50 \\

$18093-2607$ & HD 165896 & HDS2560Aa,Ab &   88937 & 
   16.4662 & 342.0 & 1.2922 &  $<$2.85\tablenotemark{a} & 692 & 40 \\
&&&&
   16.4662 & 342.0 & 1.2932 &  $<$2.48\tablenotemark{a} & 880 & 50 \\
&&&&
   17.4301 & 342.3 & 1.2830 &  $<$2.96\tablenotemark{a} & 692 & 40 \\
&&&&
   17.4301 & 342.3 & 1.2871 &  $<$2.57\tablenotemark{a} & 880 & 50 \\
&&&&
   18.2464 & 343.2 & 1.2702 &  $<$3.00\tablenotemark{a} & 692 & 40 \\
&&&&
   18.2464 & 343.2 & 1.2741 &  $<$2.51\tablenotemark{a} & 880 & 50 \\

$18099+0307$ & ADS 11113 & YSC 132Aa,Ab & 89000 &  
   15.5277 & 149.9 & 0.0209 & 0.11 & 692 & 40 \\
&&&&
   15.5277 & 149.6 & 0.0214 & 0.01 & 880 & 50 \\
&&&&
   18.2525 & 149.6 & 0.0179 & 0.22 & 562 & 44 \\
&&&&
   18.2525 & 151.0 & 0.0201 & 0.57 & 832 & 40 \\

$18112-1951$ & ADS 11127 & TOK  57Aa,Ab &   89114 &  
   17.4328 &  24.8 & 0.0119 & 1.38 & 692 & 40 \\
&&&&
   17.4328 &  15.9 & 0.0114 & 0.69 & 880 & 50 \\
$18112-1951$ & ADS 11127 & BU  132Aa,B &   89114 & 
   17.4328 & 186.6 & 1.4110 &  $<$0.79\tablenotemark{a} & 692 & 40 \\
&&&&
   17.4328 & 186.6 & 1.4154 &  $<$0.42\tablenotemark{a} & 880 & 50 \\

$18126-7340$ & HD 165259 & TOK  58Aa,Ab &   89234 &  
   17.4439 & 257.3 & 0.0855 & 3.90 & 692 & 40 \\
&&&&
   17.4439 & 255.6 & 0.0797 & 3.20 & 880 & 50 \\

$18177-1940$ & ADS 11228 & BU  246A,Ba &   89647 & 
   17.4410 & 115.4 & 0.5183 & 0.41 & 692 & 40 \\
&&&&
   17.4410 & 115.4 & 0.5178 & 0.39 & 880 & 50 \\
$18177-1940$ & ADS 11228 & WSI  89Ba,Bb &   89647 & 
   17.4410 & 336.7 & 0.0509 & 0.98 & 692 & 40 \\
&&&&
   17.4410 & 336.9 & 0.0507 & 0.97 & 880 & 50 \\

$18267-3024$ & HD 169586 & TOK 421 &   90397 &   
   16.4662 & 103.1 & 0.0123 & 2.28 & 692 & 40 \\
&&&&
   16.4662 & 103.8 & 0.0157 & 2.22 & 880 & 50 \\
&&&&
   17.4439 & 289.2 & 0.0805 & 4.42 & 692 & 40 \\
&&&&
   17.4439 & 288.2 & 0.0758 & 3.72 & 880 & 50 \\

$18340-3301$ & HR 6960 & OUD 7 &   91014 & 
   16.4662 & 177.5 & 0.2497 & 5.00 & 692 & 40 \\
&&&&
   16.4662 & 178.1 & 0.2500 & 4.56 & 880 & 50 \\
&&&&
   17.4439 & 179.0 & 0.2490 & 4.43 & 692 & 40 \\
&&&&
   17.4439 & 180.0 & 0.2461 & 3.21 & 880 & 50 \\

$18384-0312$ & HD 172088 & A 88AB &   91394 &     
   16.4880 & 193.8 & 0.1316 & -0.41\tablenotemark{d} & 562 & 40 \\
&&&&
   16.4880 & 193.3 & 0.1327 & 0.27 & 880 & 50 \\

$19264+4928$ & GJ 1237 & YSC 134 & 95575 &       
   15.5277 &  39.9\tablenotemark{b}  & 0.0199 & 0.12 & 692 & 40 \\
&&&&
   15.5277 &  49.6\tablenotemark{b} & 0.0242 & 0.64 & 880 & 50 \\

$19401-0759$ & HD 185588 & YSC 161 &   96754 &  
   17.4464 & 252.4 & 0.0710 & 0.85 & 692 & 40 \\
&&&&
   17.4464 & 252.0 & 0.0706 & 0.77 & 880 & 50 \\

$19453-6823$ & HD 185655 & HDS2806A,Ba &   97196 & 
   17.4441 & 134.4 & 0.3542 & 3.03 & 692 & 40 \\
&&&&
   17.4441 & 134.5 & 0.3534 & 2.26 & 880 & 50 \\
$19453-6823$ &HD 185655  & TOK 425Ba,Bb &   97196 & 
   17.4441 & 155.3 & 0.0403 & 0.60 & 692 & 40 \\
&&&&
   17.4441 & 147.6 & 0.0358 & 1.12 & 880 & 50 \\

$20048+0109$ & HD 190412 & TOK 699 &   98878 &  
   17.4386 &  46.2 & 0.1261 & 2.97 & 692 & 40 \\
&&&&
   17.4386 &  46.4 & 0.1264 & 2.40 & 880 & 50 \\

$20275-0206$ & ADS 13868 & LSC 89Aa,Ab &  100896 & 
   15.5280 & 202.2 & 0.0151 & 0.31 & 692 & 40 \\
&&&&
   15.5280 & 196.7 & 0.0160 & 0.54 & 880 & 50 \\

$20393-1457$ &ADS 14099  & HU 200AB &  101923 &  
   17.4414 & 120.4 & 0.3207 & 0.47 & 692 & 40 \\
&&&&
   17.4414 & 120.4 & 0.3205 & 0.67 & 880 & 50 \\

$21041+0300$ & HD 200580 & WSI 6AB & 103987 &      
   15.5280 & 287.0 & 0.2474 & 1.83 & 692 & 40 \\
&&&&
   15.5280 & 287.1 & 0.2445 & 1.31 & 880 & 50 \\
&&&&
   17.4414 & 302.0 & 0.1996 & 2.03 & 692 & 40 \\
&&&&
   17.4414 & 302.0 & 0.2004 & 1.44 & 880 & 50 \\

$22300+0426$ & ADS 15988 & STF2912A,Ba &  111062 & 
   17.4388 & 296.5 & 0.1112 & 1.93 & 692 & 40 \\
&&&&
   17.4388 & 297.0 & 0.1106 & 1.88 & 880 & 50 \\
$22300+0426$ & ADS 15988 & STF2912Ba,Bb &  111062 & 
   17.4388 & 297.4 & 0.0313 & 0.31 & 692 & 40 \\
&&&&
   17.4388 & 295.2 & 0.0324 & 0.35 & 880 & 50 \\

$22357+5312$ & HD 214222 & A 1470 & 111528 &       
   15.5363 &  66.3 & 0.1121 & 0.43 & 692 & 40 \\
&&&&
   15.5363 &  66.5 & 0.1109 & 0.29 & 880 & 50 \\

$22388+4419$ & HD 214608 & HO 295A,Ba & 111805 & 
   15.5446 & 333.9 & 0.3324 & 0.00 & 692 & 40 \\
&&&&
   15.5446 & 334.0 & 0.3304 & 0.64 & 880 & 50 \\
$22388+4419$ & HD 214608 & BAG 15Ba,Bb & 111805 & 
   15.5446 & 151.7 & 0.0143 & 0.62 & 692 & 40 \\
&&&&
   15.5446 & 147.7 & 0.0150 & -0.55\tablenotemark{d} & 880 & 50 \\

$23347+3748$ & HD 221757 & YSC 139 & 116360 &   
   15.5446 &  92.5 & 0.0305 & 0.58 & 692 & 40 \\
&&&&
   15.5446 &  92.8 & 0.0300 & 0.53 & 880 & 50 \\

$23485+2539$ & HD 223323 & DSG 8 & 117415 &     
   15.5447 & 129.2 & 0.0255 & 0.13 & 692 & 40 \\
&&&&
   15.5447 & 130.4 & 0.0255 & 0.24 & 880 & 50 \\

\enddata 
 
\tablenotetext{a}{Photometry for this observation appears as an upper limit because
the observation may be affected by speckle decorrelation as discussed in the text.}
\tablenotetext{b}{Quadrant inconsistent with previous measures in the 4th Interferometric Catalog (Hartkopf \ea 2001a).}
\tablenotetext{c}{A magnitude difference for this single-lined spectroscopic binary does not
appear for the reasons discussed in the text for small-separation systems.}
\tablenotetext{d}{ The position angle of this observation was found to be in the opposite quadrant than
that of the observation in the other filter. To keep the position angles consistent, we adopt a negative
magnitude difference for this observation.}
\tablenotetext{e}{ The component was undetected in this filter, with a lower limit in the magnitude difference shown.}

\end{deluxetable} 

\clearpage

\begin{deluxetable}{clrcrrrrl}
\tabletypesize{\tiny}
\tablewidth{0pt}
\tablenum{3}
\tablecaption{Ephemeris Positions and Residuals Used in the Astrometric Accuracy Study}
\tablehead{
\colhead{WDS} &
\colhead{Discoverer} &
\colhead{HIP} &
\colhead{Date} &
\colhead{$\theta_{eph}$} &
\colhead{$\rho_{eph}$} &
\colhead{${\Delta \bar{\theta}}$} &
\colhead{${\Delta \bar{\rho}}$} &
\colhead{WDS Orbit Grade} \\
& \colhead{Designation} &&
\colhead{($2000+$)} &
\colhead{($^{\circ}$)} &
\colhead{($^{\prime \prime}$)} &
\colhead{($^{\circ}$)} &
\colhead{(mas)} & \colhead{and Reference}
}
\startdata
$02396-1152$ & FIN 312 &  12390 & 17.9477 &
  $200.9 \pm 0.4$ & $0.0855 \pm 0.0004$ & $+2.4$ & $-2.5$ & 1, Docobo \& Andrade (2013) \\

$04107-0452$ & A 2801 & 19508 & 16.0368 & 
  $345.7 \pm 0.7$ & $0.1304 \pm 0.0017$ & $+0.5$ & $-2.9$ & 2, Tokovinin (2017) \\
&&& 16.0476 &
  $345.8   \pm  0.7$  & $0.1307 \pm  0.0017$ & $+0.6$ & $-2.1$ & \\
&&& 18.2460 &
  $355.0   \pm  0.8$  & $0.1658 \pm  0.0023$ & $+0.7$ & $-2.9$ & \\

$04136+0743$ & A 1938 & 19719 & 17.9477 &
 $328.1 \pm 0.1$ &  $0.1377 \pm 0.0002$ & $+1.9$ & $+2.0$ & 1, Muterspaugh \ea (2010) \\
  
$16044-1122$ & STF 1998 &78727 & 18.2490 &
  $9.0 \pm 5.9$ & $1.1145 \pm 0.0147$ & $-0.0$ & $+8.2$ & 1, Docobo \& Ling (2009) \\

$18099+0307$ &  YSC 132Aa,Ab & 89000 & 15.5277 &
  $149.6 \pm 2.7$ & $0.0206 \pm 0.0006$ & $+0.1$ & $+0.6$ & 2, M\'{e}ndez \ea (2017) \\
&&& 18.2525 &
  $152.9 \pm 2.8$ & $0.0207 \pm 0.0006$ & $-2.6$ & $-1.7$ & \\

$18384-0312$ & A 88AB & 91394 & 16.4880 &
  $193.4 \pm 2.9$ & $0.1351 \pm 0.0026$ & $+0.1$ & $-3.0$ & 1, Griffin (2013)\\


\enddata


\end{deluxetable}

\begin{deluxetable}{clrcrrrr}
\tabletypesize{\scriptsize}
\tablewidth{0pt}
\tablenum{4}
\tablecaption{A Comparison of Astrometry from Table 2 with {\it Gaia} DR2 results}
\tablehead{
\colhead{WDS} &
\colhead{Discoverer} &
\colhead{HIP} &
\colhead{Date} &
\colhead{$\theta_{Gaia}$} &
\colhead{$\rho_{Gaia}$} &
\colhead{${\Delta \bar{\theta}}$} &
\colhead{${\Delta \bar{\rho}}$}  \\
& \colhead{Designation} &&
\colhead{($2000+$)} &
\colhead{($^{\circ}$)} &
\colhead{($^{\prime \prime}$)} &
\colhead{($^{\circ}$)} &
\colhead{(mas)}
}
\startdata


$16120-1928$ & BU 120Aa-B &  79374  & 17.4298  &  2.4  & 1.3263  &  $-$1.4 & $+$8.0 \\
$17066+0039$ & BU 823A,Ba & 83716   & 17.4327  & 173.6 &  0.9901 & $-$3.8 & $+$20.6 \\
$18093-2607$ & HDS2560Aa,Ab & 88937  & 16.4662  &  340.2  & 1.2988 & $+$1.8 & $-$6.1 \\
  &&                                                    88937  & 17.4301  &  340.2  & 1.2988 & $+$2.1 & $-$13.8 \\
 &&                                                     88937   &18.2464   &  340.2  & 1.2988 & $+$3.0 & $-$26.7 \\
$18112-1951$ & BU  132Aa,B &  89114  & 17.4328  &  184.5  & 1.4051 & $+$2.1 & $+$8.1 \\
\enddata

\end{deluxetable}

\clearpage

\begin{deluxetable}{lccccc}
\tabletypesize{\tiny}
\tablewidth{0pt}
\tablenum{5}
\tablecaption{Orbital Elements for Five Systems}
\tablehead{ 
\colhead{Parameter} &
\colhead{WCK 1Aa,Ab} &
\colhead{YSC 129} &
\colhead{LSC 45} &
\colhead{DSG 7Aa,Ab} &
\colhead{DSG 8} 
} 
\startdata 


HIP & 5336 & 32040 & 36387 & 81023 & 117415  \\
Sp. Orbit Type & SB1 & SB2 & none & SB1 & SB2\\
$P$, years       & $21.479 \pm 0.032$ & $2.5594 \pm 0.0035$ & $40.8 \pm 2.6$  & $0.61867 \pm 0.00033$ & $3.2191 \pm 0.0038$ \\
$a$, mas           & $989.8 \pm 5.4$ & $33.7 \pm 0.7$ & $95.3 \pm 3.3$ & $16.0 \pm 1.2$ & $40.8 \pm 1.0$  \\
$i$, degrees       & $111.40 \pm 0.27$ & $105.2 \pm 2.5$ & $59.2 \pm 1.0$ & $117.9 \pm 8.2$  & $86.0 \pm 0.9$  \\
$\Omega$, degrees  & $41.28 \pm 0.63$ & $280.7 \pm 1.4$ & $ 24.6 \pm 63.6$ & $162.7 \pm 8.2$ & $ 121.1 \pm 1.1$  \\
$T_{0}$, years    & $1975.5514 \pm 0.0805$ & $2002.7371 \pm 0.0155$ & $2011.6 \pm 8.0$  & $1988.4316 \pm 0.0030$ & $2011.0545 \pm 0.0085$  \\ 
$e$               & $0.6251 \pm 0.0099$ & $0.158 \pm 0.007$ & $0.092 \pm 0.033$  & $0.3067 \pm 0.0088$ & $0.602 \pm 0.008$  \\
$\omega$, degrees & $144.93 \pm 1.63$ & $296.8 \pm 2.3$ & $ 193.9 \pm 34.9$ & $20.46 \pm 1.74$ & $ 78.4 \pm 0.6$ \\
$K_{1}$ (km/s) & $2.62 \pm 0.11$ & $14.35 \pm 0.13$ & ... & $23.62 \pm 0.23$ & $16.41 \pm 0.18$ \\
$K_{2}$ (km/s) & ... & $15.07 \pm 0.13$ & ... & ... & $16.24 \pm 0.22$ \\
$v_{0}$ (km/s) & $-97.560 \pm 0.063$ & $82.65 \pm 0.07$  & ... & $-51.28 \pm 0.15$ & $ -9.50 \pm 0.07$\\

\enddata 


\end{deluxetable}

\begin{deluxetable}{lrrccrcc}
\tabletypesize{\scriptsize}
\tablewidth{0pt}
\tablenum{6}
\tablecaption{Further Observed Properties for the Systems in Tables 5.}
\tablehead{ 
\colhead{Name} &
\colhead{HIP} &
\colhead{$\pi$} & 
\colhead{$M_{V}$} & 
\colhead{Spectral} & 
\colhead{[Fe/H]} &
\colhead{$B-V$} &
\colhead{$\Delta m$} \\
&&
\colhead{(mas)} & 
\colhead{(mag.)} & 
\colhead{Type} & 
&& (692 nm)
} 
\startdata 
WCK 1Aa,Ab & 5336 & $132.38 \pm 0.82$\tablenotemark{a} & 5.78 & G5V & $-0.68$\tablenotemark{b} & $0.704 \pm 0.004$ & $5.70 \pm 0.74$ \\
YSC 129 & 32040 & $14.4824 \pm 0.0705$ & 2.29 & F8III & $-0.20$ & $0.509 \pm 0.006$ & $0.64 \pm 0.11$ \\
LSC 45 & 36387 & $6.5387 \pm 0.0614$ & 2.09 & F2 & $-0.52$ & $0.438 \pm 0.014$ & $1.40 \pm 0.02$ \\
DSG 7Aa,Ab    &  81023 & $24.0090 \pm 0.3582$ & 5.64 & K0V  & 
$-1.39$\tablenotemark{c}  & $0.868 \pm 0.004$ & $1.13 \pm 0.16$ \\
DSG 8         & 117415 & $14.3551 \pm 0.0532$ & 2.16 & F2IV-V & $-0.46$ & $0.443 \pm 0.009$ & $0.07 \pm 0.02$ \\
\enddata 
\tablenotetext{a}{No {\it Gaia} DR2 parallax is available; the value shown is the revised 
{\it Hipparcos} result.}
\tablenotetext{b}{From Boyajian \ea (2008).}
\tablenotetext{c}{From Latham \ea (1992).}
\end{deluxetable}

\begin{deluxetable}{lrlcclccc}
\tabletypesize{\scriptsize}
\tablewidth{0pt}
\tablenum{7}
\tablecaption{Mass Comparison for the Systems in Table 6.}
\tablehead{ 
\colhead{Name} &
\colhead{HIP} &
\colhead{Assigned Component} &
\colhead{Derived} &
\colhead{Derived} &
\colhead{Derived} &
\colhead{Implied Total} &
\colhead{Total Mass} \\
&&
\colhead{Spectral Types} &
\colhead{Composite} &
\colhead{$\Delta m$} &
\colhead{Masses} &
\colhead{Mass} &
\colhead{from Orbit } \\
&&      and $B-V$ Colors\tablenotemark{a} & $B-V$\tablenotemark{a} &
\colhead{at 692 nm} &
\colhead{($M_{\odot}$)\tablenotemark{a}} &
\colhead{($M_{\odot}$)\tablenotemark{a}} &
\colhead{($M_{\odot}$)}

} 
\startdata 
WCK 1Aa,Ab & 5336 & G5V,M4.5V (0.68,1.59) & 0.68 & 5.62 & 0.92, 0.24 & $1.16 \pm 0.07$ & $0.906 \pm 0.023$ \\
YSC 129 & 32040 & F5IV,G5IV (0.44,0.68) & 0.49 & 0.43 & 1.4, 1.0 & $2.40 \pm 0.14$ & $1.83 \pm 0.14$ \\
LSC 45 & 36387 &  F2V,G2V (0.35,0.63) & 0.41 & 1.40 & 1.52, 1.02  & $2.54 \pm 0.05$ & $1.86 \pm 0.31$ \\
DSG 7Aa,Ab    &  81023 & K1V,K6V (0.86,1.24) & 0.88 & 1.30 & 0.77, 0.64 & $1.41 \pm 0.03$ & $0.70 \pm 0.20$  \\
DSG 8         & 117415 & F5V,F5.5V (0.44,0.45) & 0.44 & 0.06 &1.40, 1.37 & $2.77 \pm 0.09$ & $2.16 \pm 0.17$ \\
\enddata 

\tablenotetext{a}{These columns assume the Solar metal abundance.}
\end{deluxetable}

\end{centering}

\clearpage

\begin{figure}[tb]
\plottwo{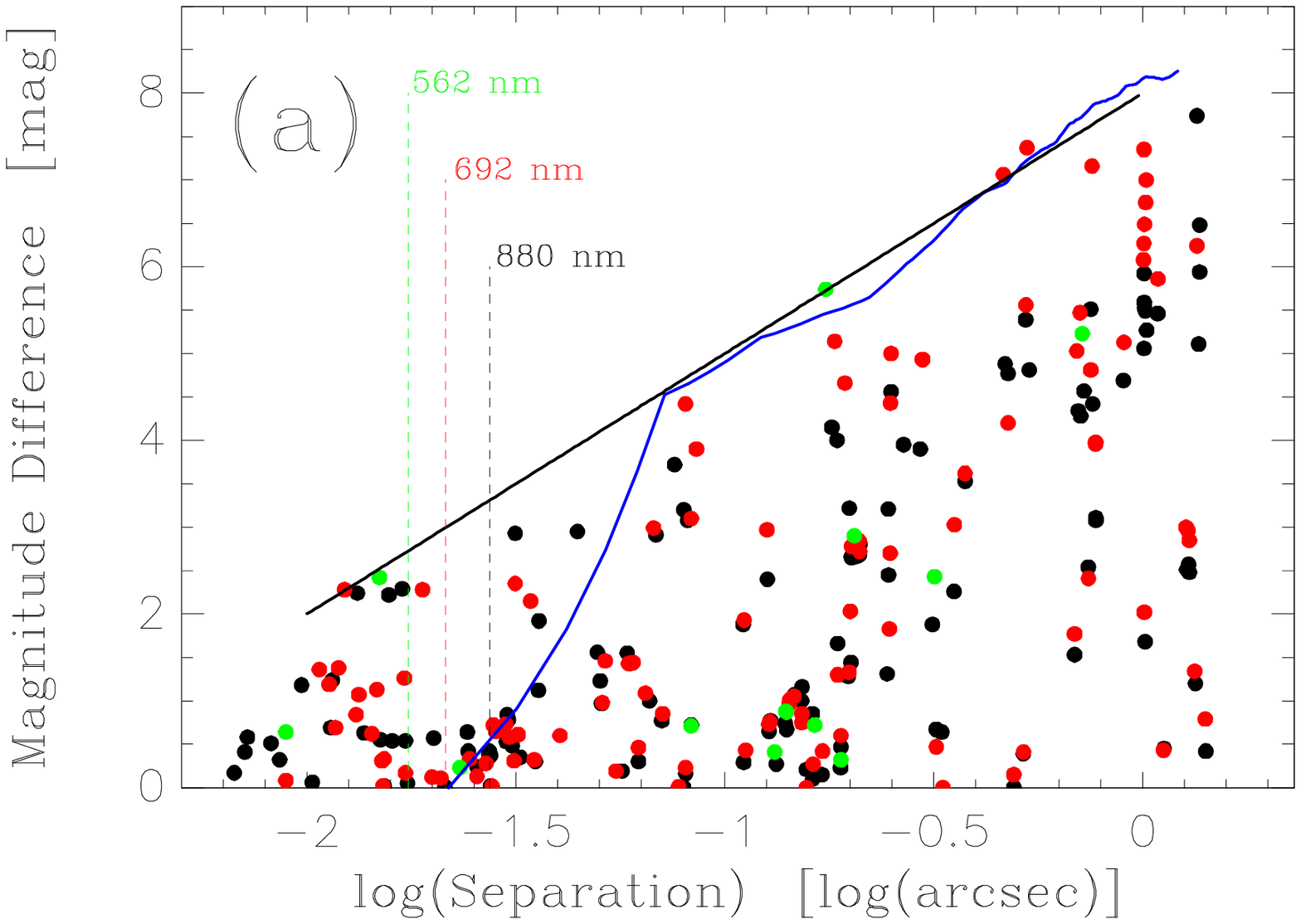}{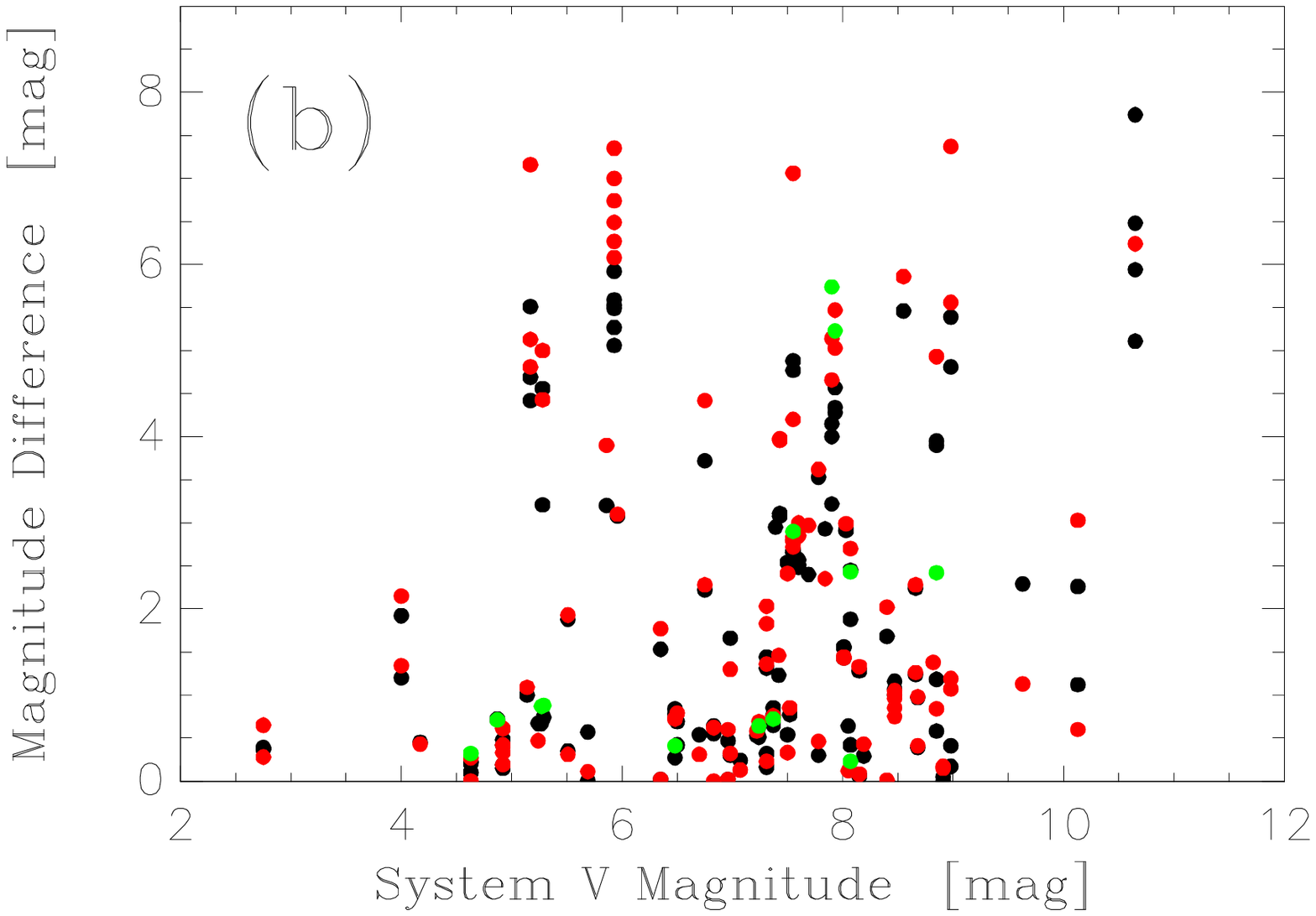}
\caption{
Magnitude difference versus (a) separation and
(b) system V magnitude for the measures appearing in Table 2. The color of the plot
symbol indicates the center wavelength of the filter used for the measure, with green indicating 562 nm, 
red indicating 692 nm and black used for both 832 or 880 nm. The blue curve in 
(a) represents a typical detection limit curve for speckle observations at Gemini,
which is measured from the noise properties of the reconstructed image and assumes 
no sensitivity to companions below the diffraction limit. The black line extends 
the measured sensitivity to the sub--diffraction-limited regime. The dashed vertical
lines indicate the location of the diffraction limit for the wavelengths indicated.}
\end{figure}

\begin{figure}[tb]
\plotone{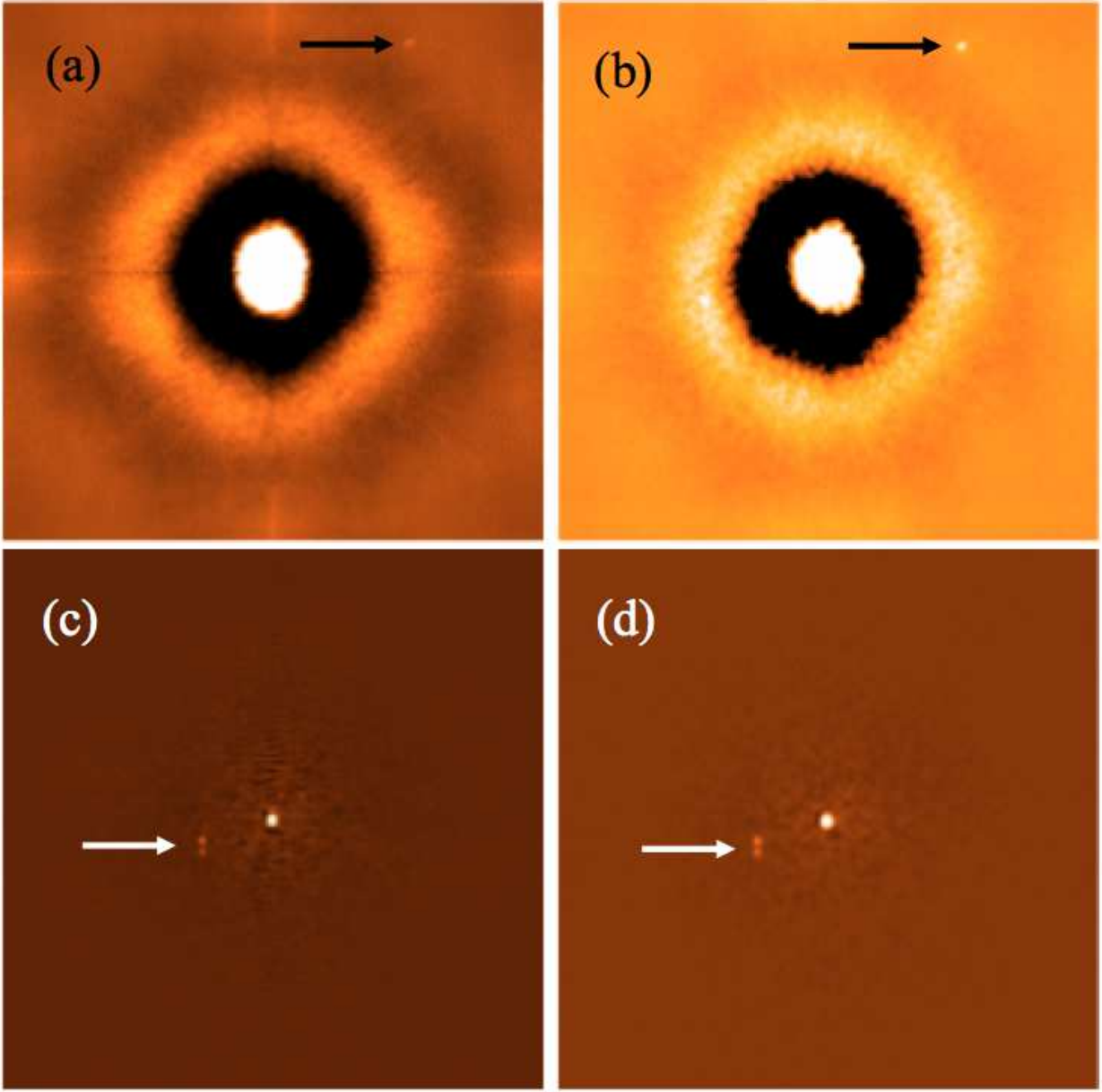}
\caption{
Reconstructed images of two systems presented in Table 2.
(a) DSG 9 = HIP 4754, at 692 nm. (b) DSG 9 at 880 nm. A very faint previously unknown companion 
(shown with the arrow in each image) was
detected to this known spectroscopic binary star (which is unresolved here). The apparent very faint peak 
to the left and slightly below the primary in the 880-nm image is a known reflection in the instrument.
(c) The triple system HD 2092 = HIP 72492 at 692 nm, (d) HDS 2092 at 880 nm. In this
case the secondary of the known binary HDS 2092 was resolved into a small-separation
pair. In all cases, north is up and east is to the right. Each image is $2.8 \times 2.8$ arcsec in size.}
\end{figure}

\begin{figure}[tb]
\plottwo{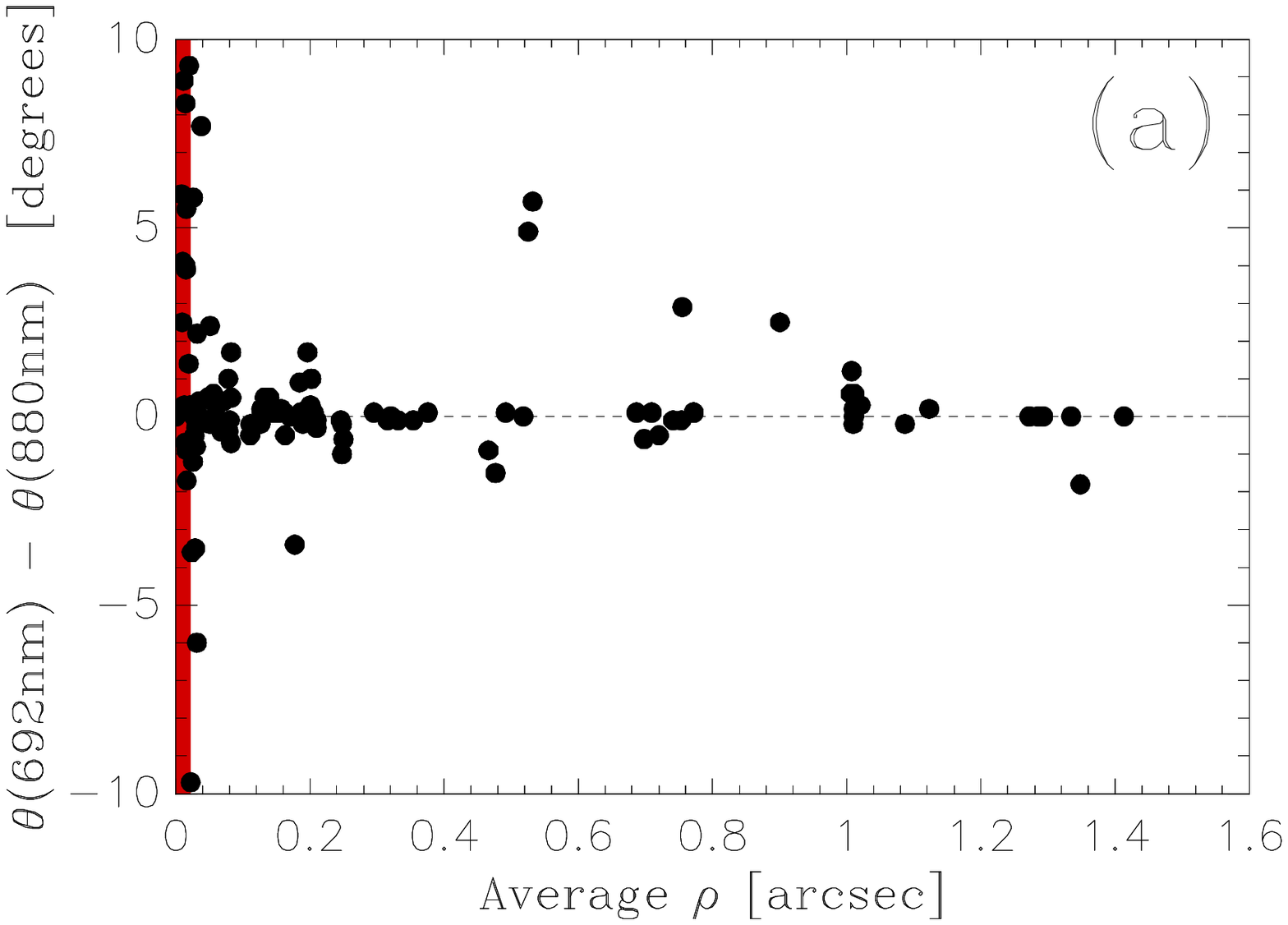}{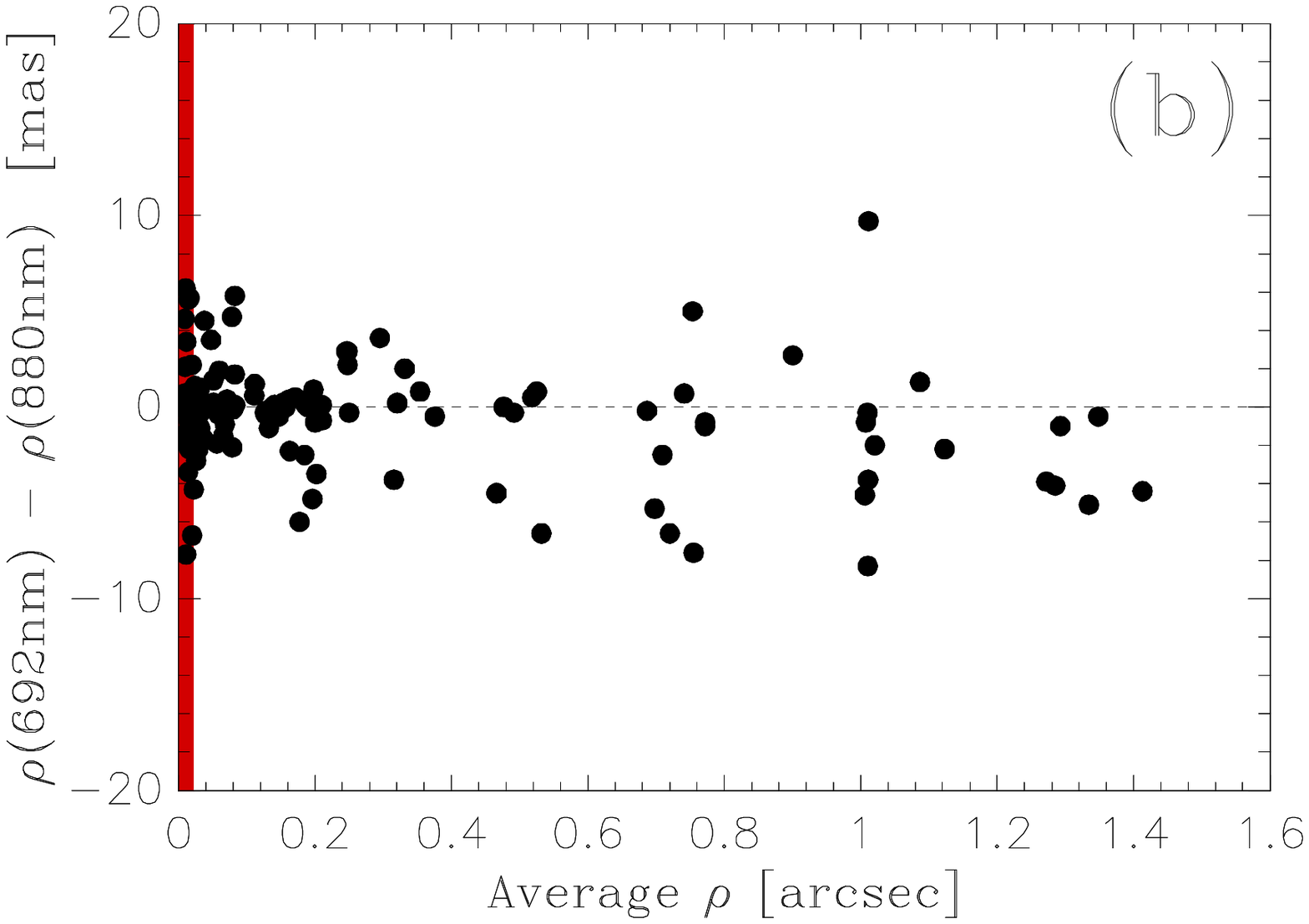}

\vspace{1.0cm}

\plottwo{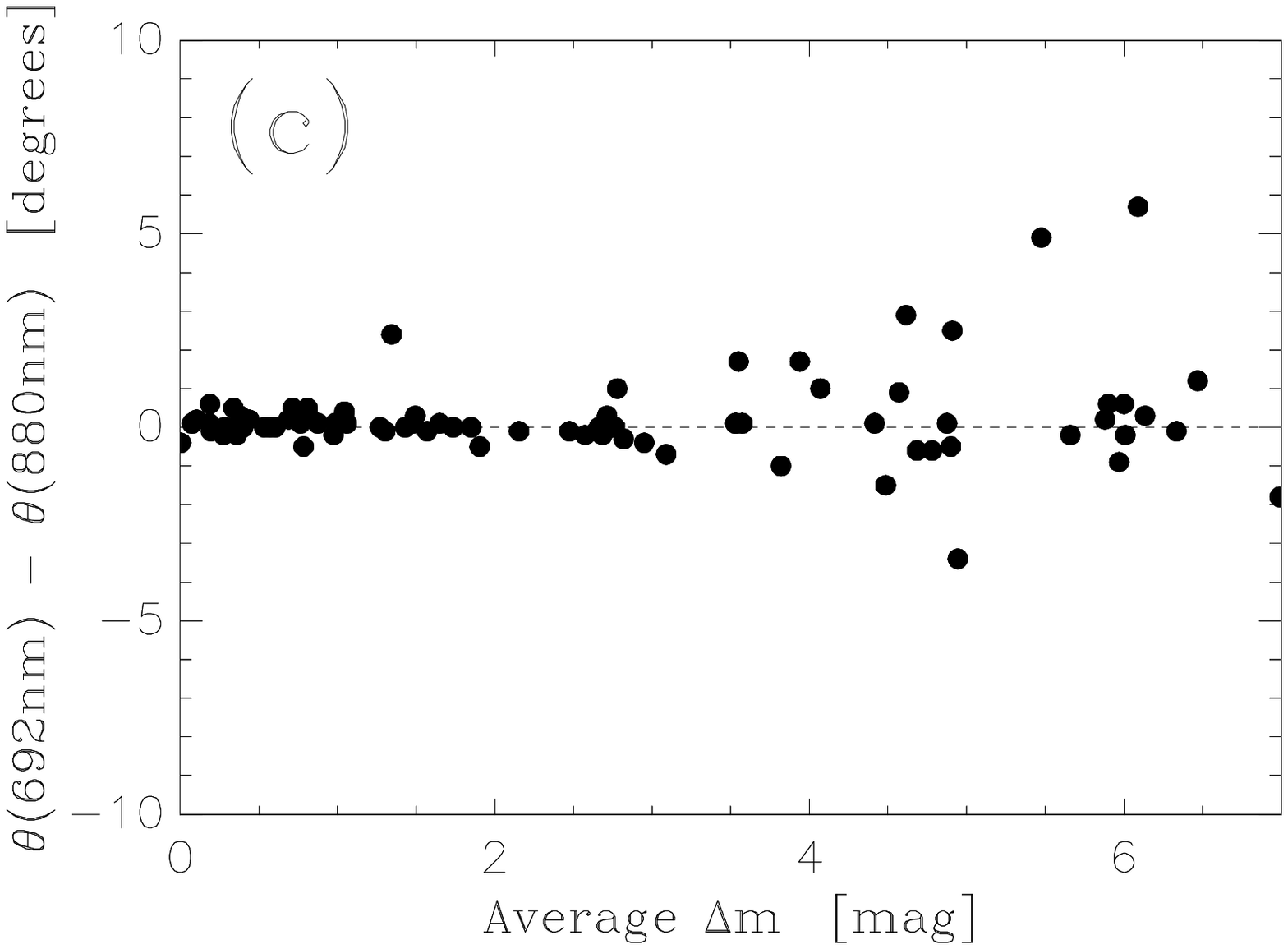}{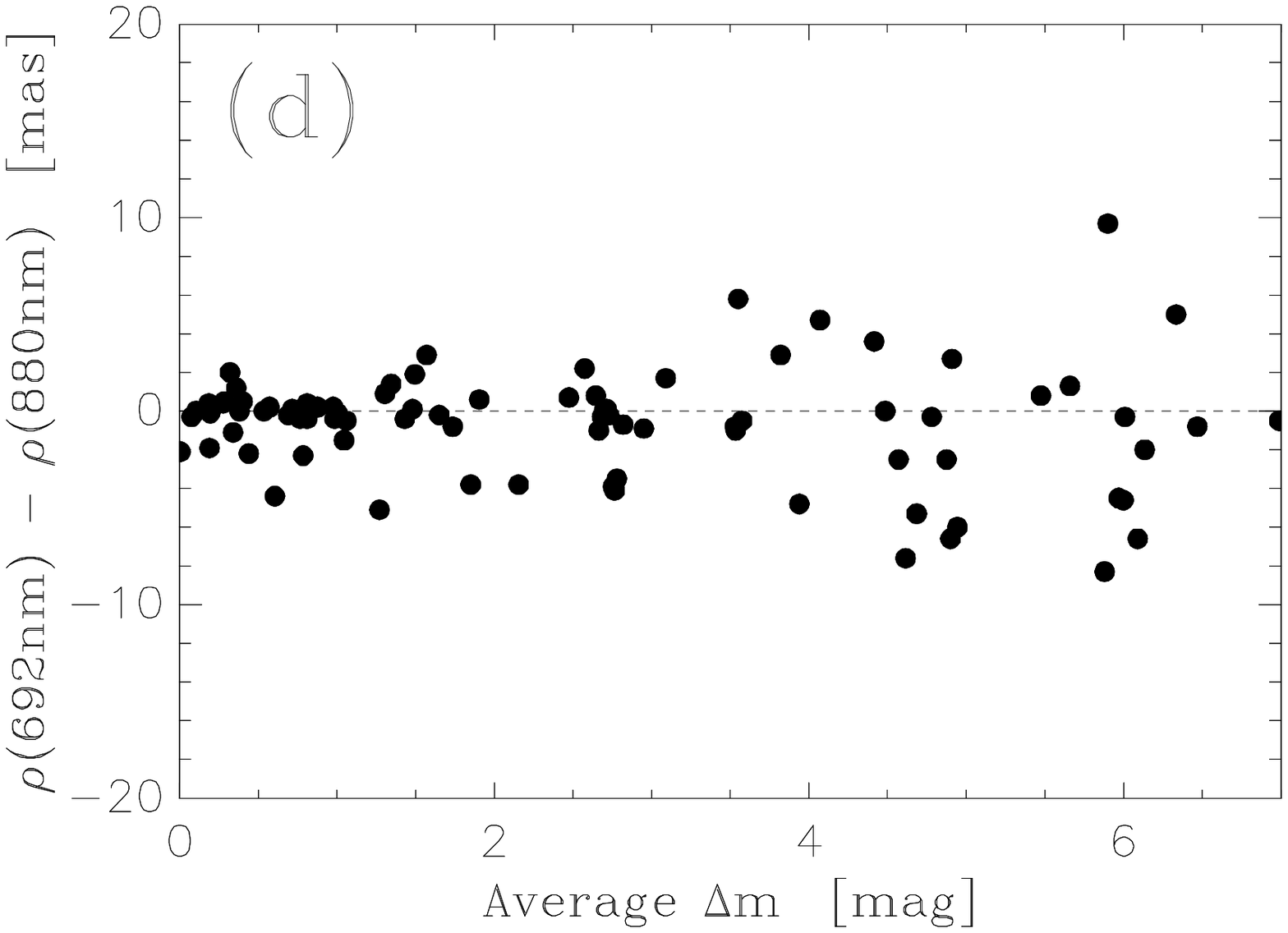}
\caption{
Measurement differences between the two channels of the instrument for
position angle and separation 
plotted as a functions of measured separation, $\rho$ in (a) and (b), and
as functions of $\Delta m$ in (c) and (d).
In (a) and (b), the dark red band at the left marks the region below the
diffraction limit of the telescope, and in all plots a dotted line is drawn at zero
to indicate perfect agreement between the channels.
}
\end{figure}

\begin{figure}[tb]
\plottwo{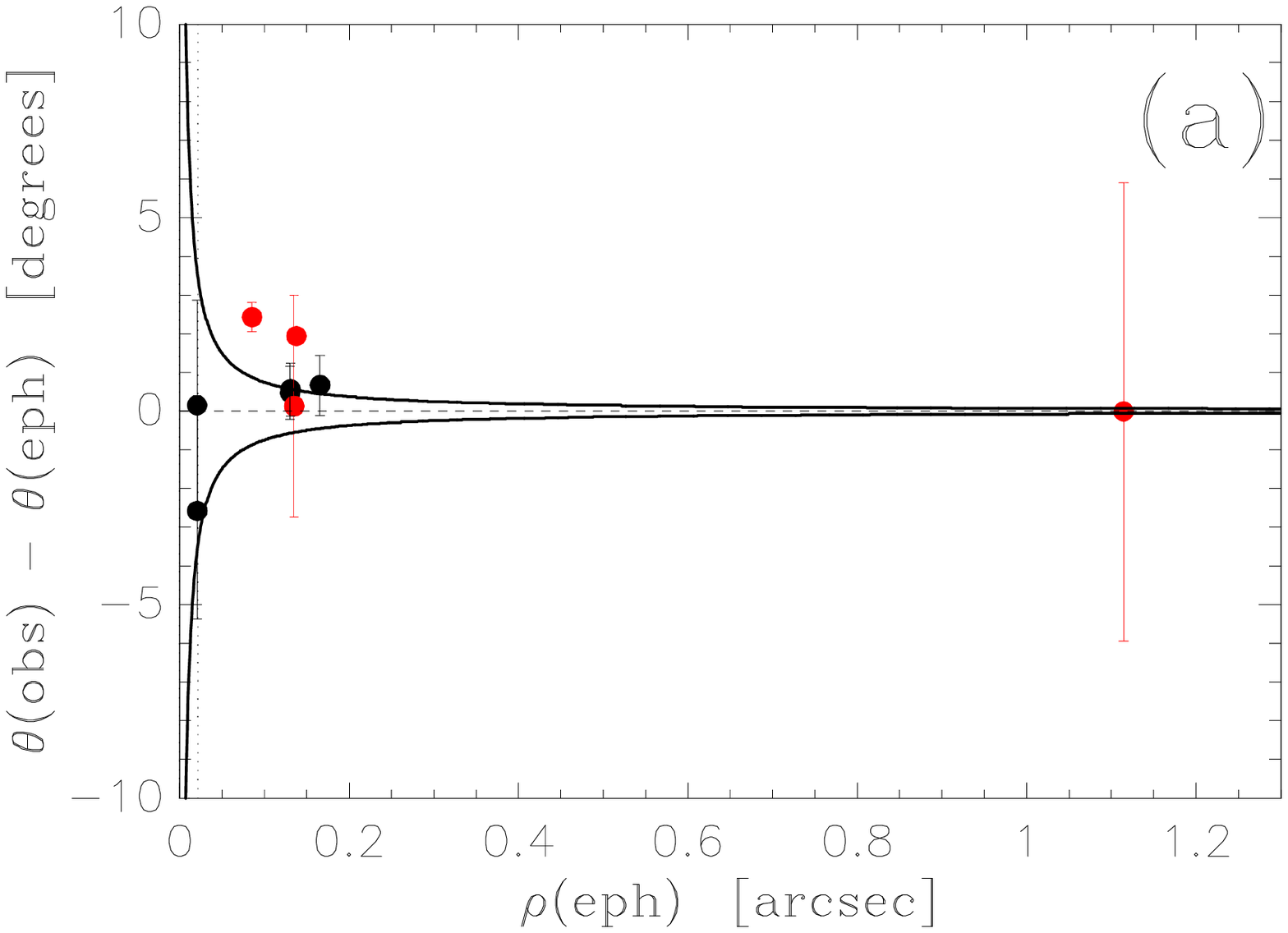}{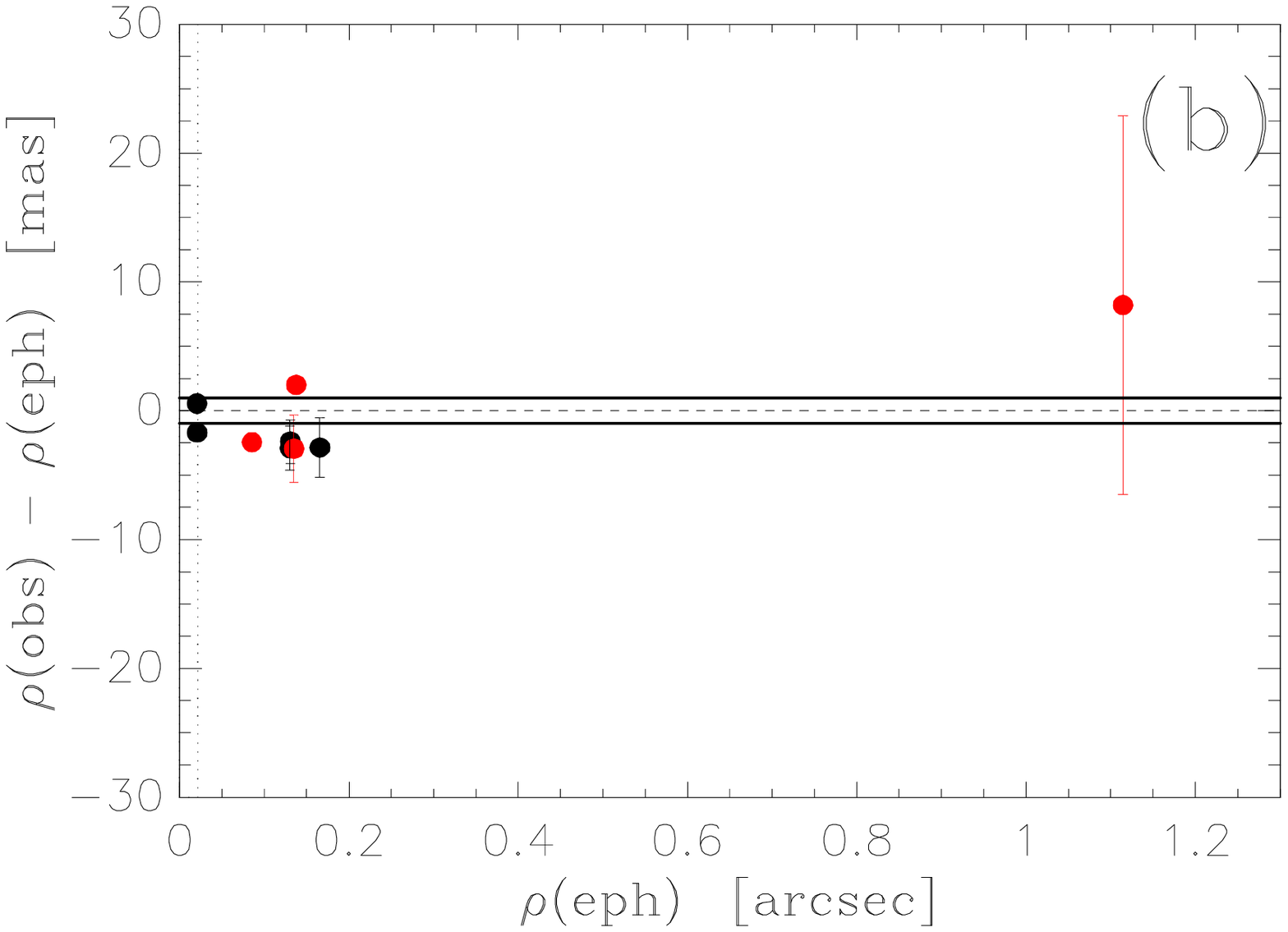}
\caption{
(a) Position angle residuals for measures in Table 2 when comparing to orbits of 
Grade 1 or 2 in the 6th Orbit Catalog (Hartkopf \ea 2001b). (b) Separation residuals for the same
observations. Red data points indicate Grade 1 orbits, and black data points are
used for Grade 2 orbits. A dashed horizontal line is drawn to indicate the zero
residual line, and in (a) a 1.3-mas uncertainty is used to create the solid curves
to indicate how that uncertainty translates into a position angle uncertainty. In
(b), solid lines are drawn at $\pm1.3$ mas to indicate the uncertainty of our
speckle measures. The vertical dotted lines mark the diffraction limit of the telescope
at 692 nm.}
\end{figure}

\begin{figure}[tb]
\plotone{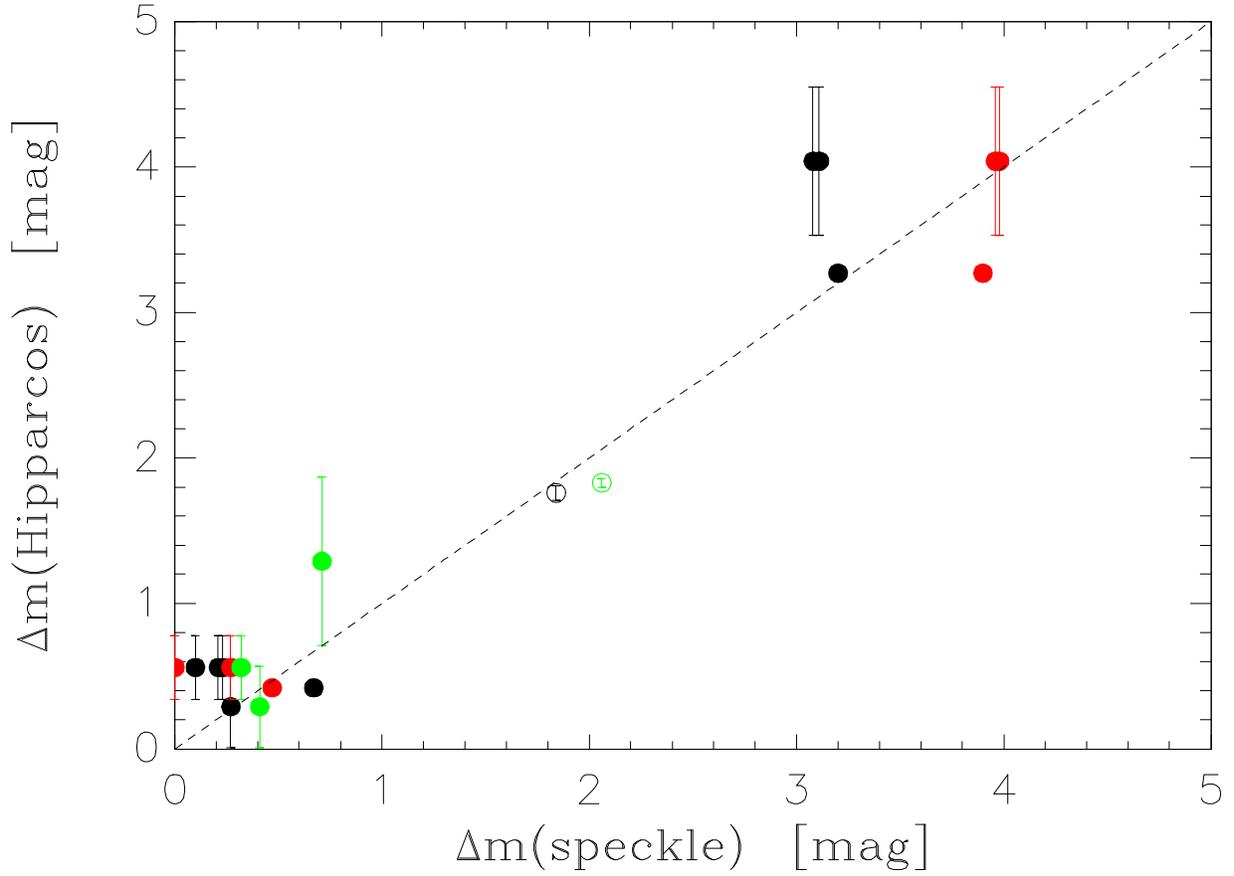}
\caption{
Measures of magnitude difference obtained in the analysis here versus 
those appearing in the {\it Hipparcos} Catalogue (using the $H_{p}$ filter). 
The open circles are the result
for HIP 75695 = JEF 1, where there is no {\it Hipparcos} measure but there
are several recent measures from our work at WIYN and the DCT, and
so those are used for the comparison here. The color of the plot symbol
indicates the filter of the speckle observation: green represents 562 nm, 
red represents 692 nm, and black represents 832 or 880 nm.
}
\end{figure}

\begin{figure}[tb]
\plottwo{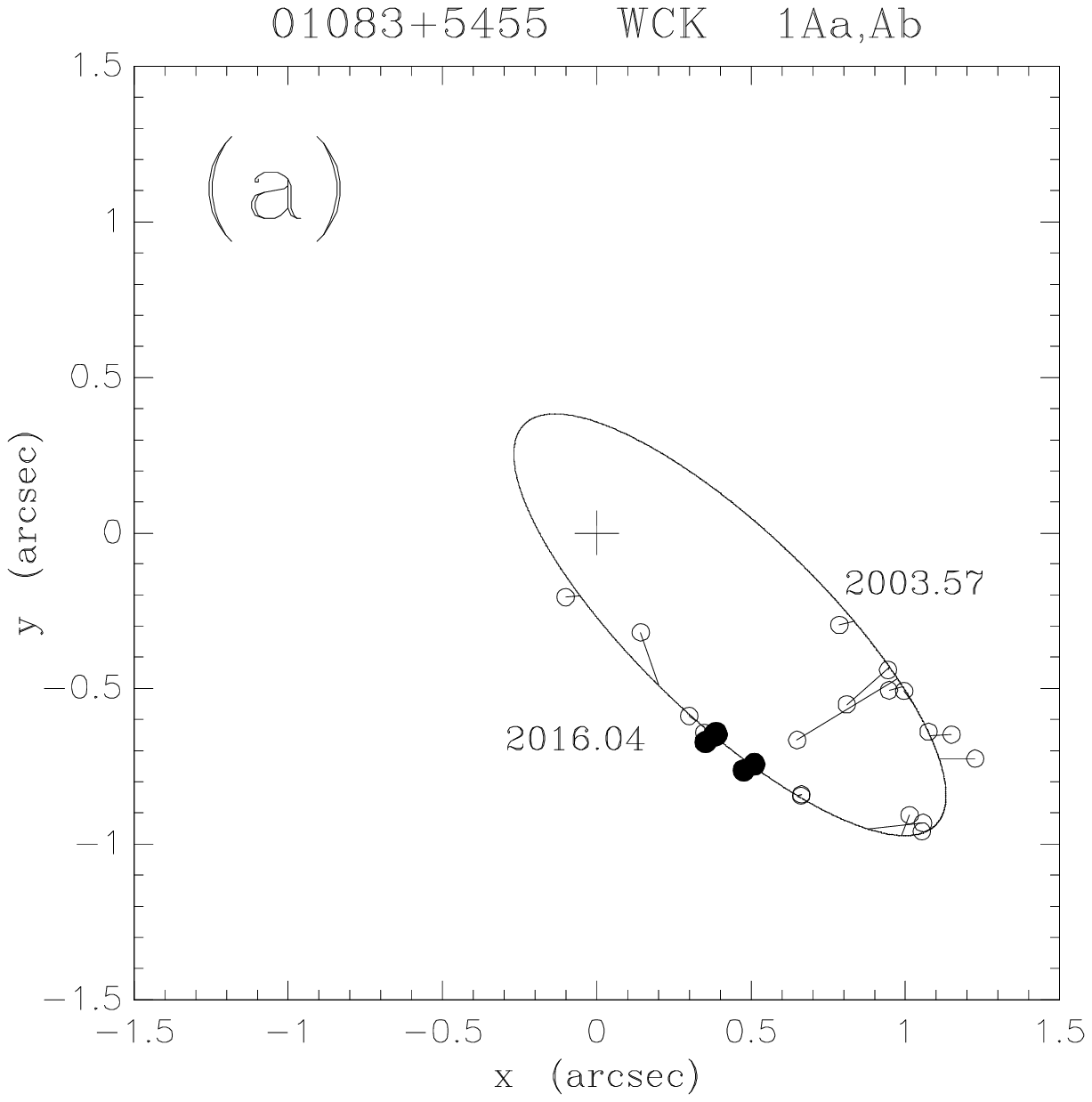}{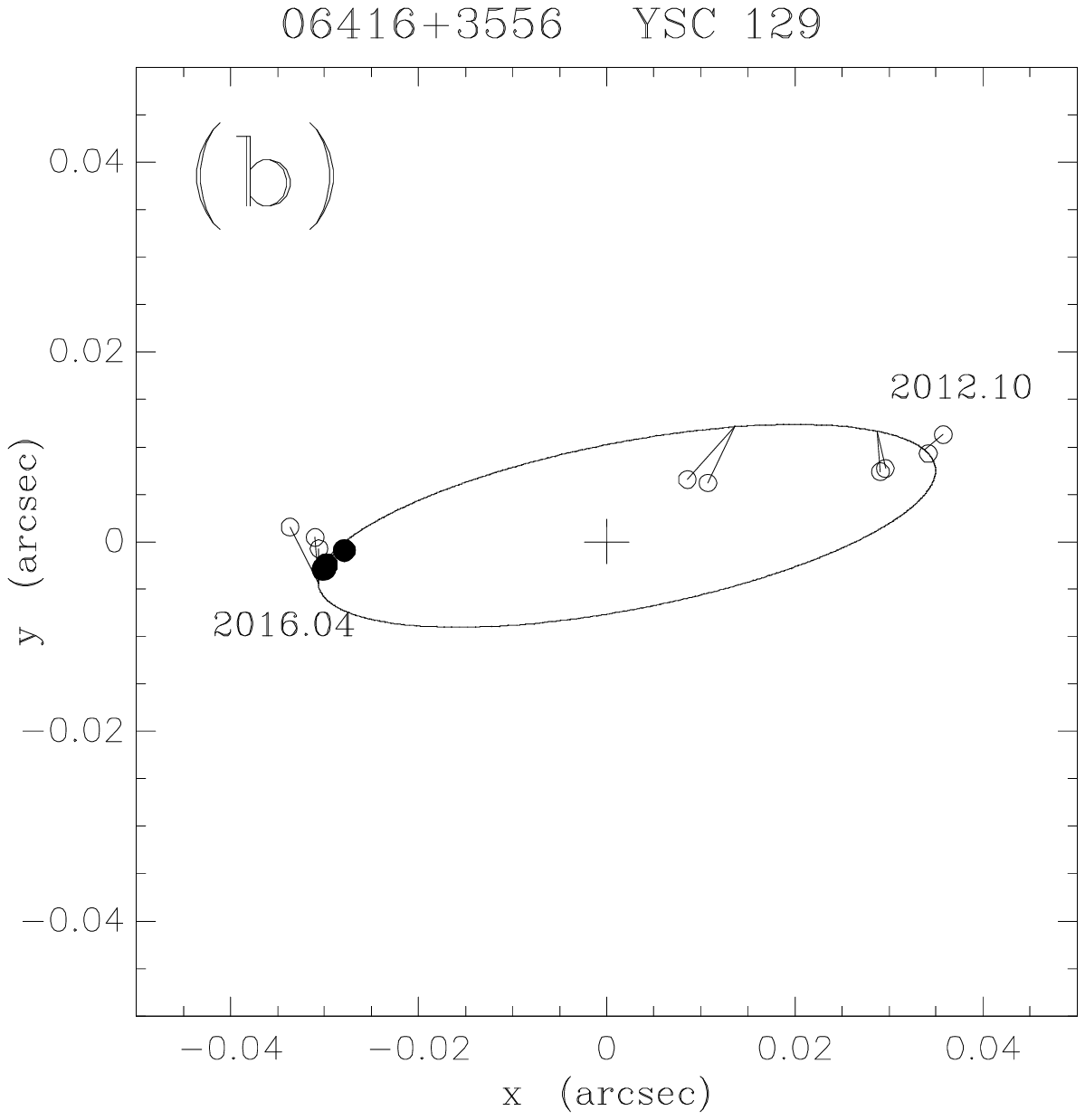}

\vspace{0.1cm}

\plottwo{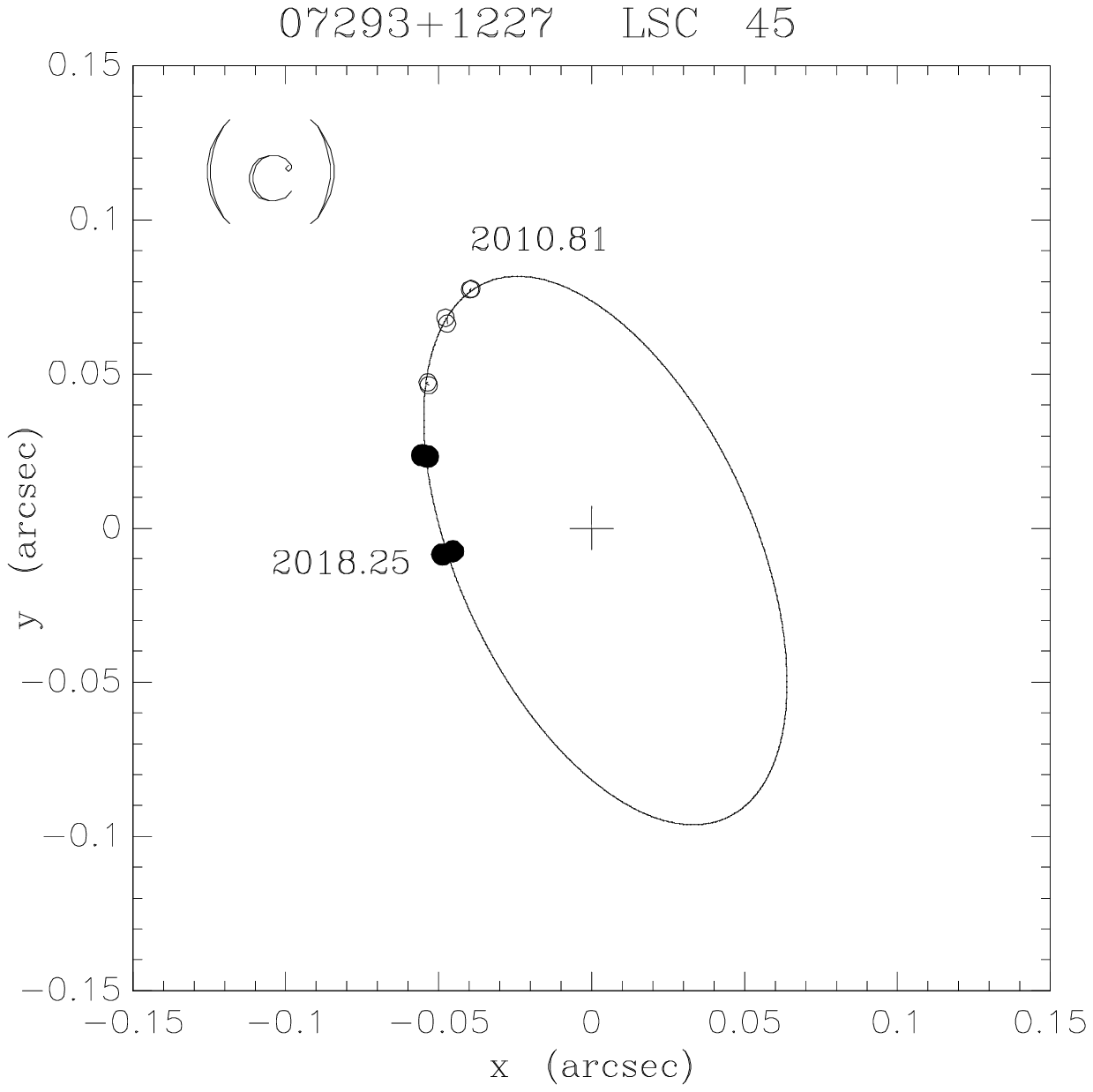}{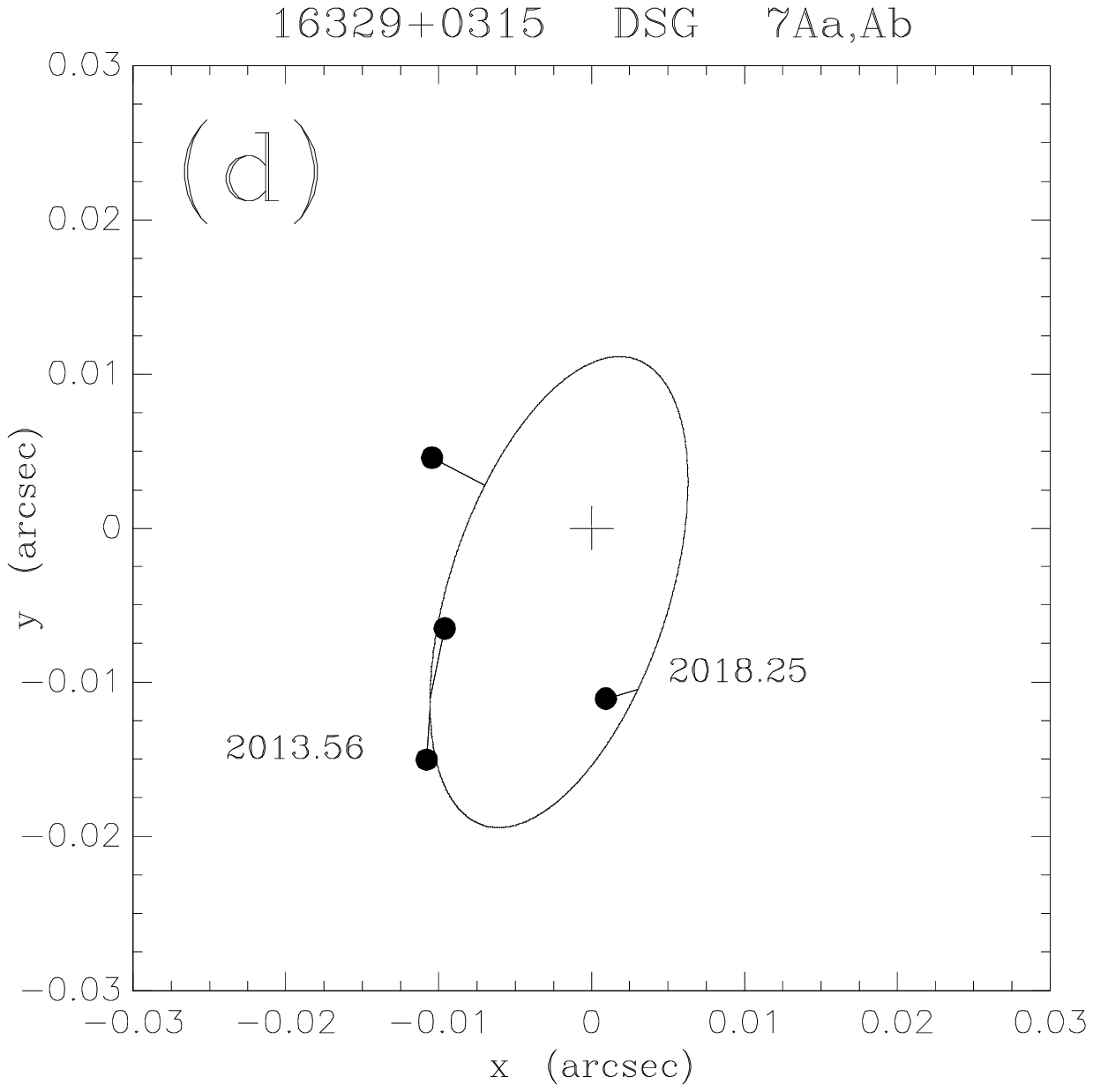}

\vspace{0.1cm}

\epsscale{0.5}
\plotone{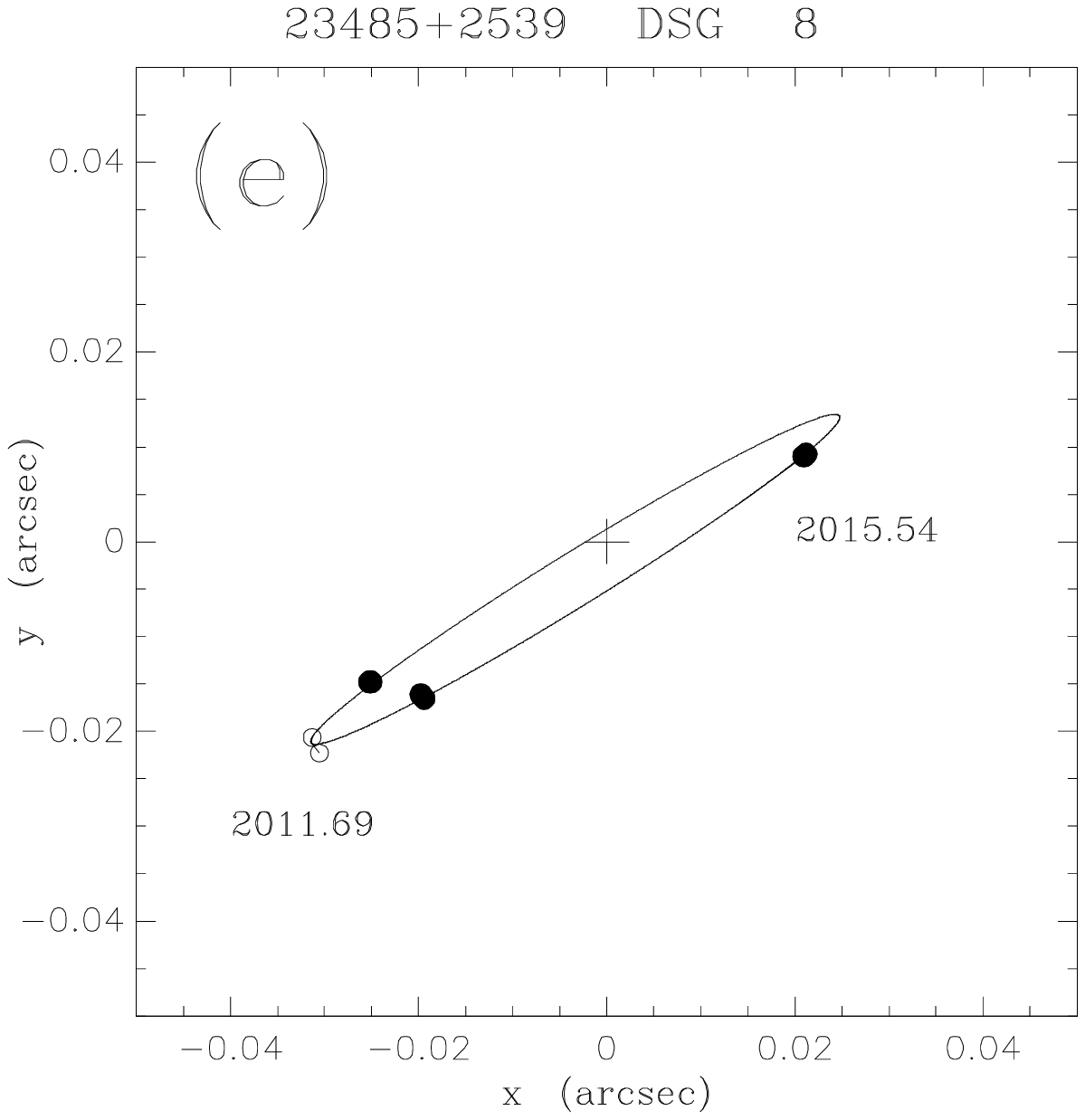}
\caption{
Visual orbits for the five systems appearing in Table 5 and described 
in the text. In all cases, the 
calculated orbit is shown as a solid line, non-Gemini observations in the 
literature are shown as open circles, and Gemini observations, including
those presented in this paper, are shown as filled circles. A line segment
is drawn from each observation to the ephemeris prediction on the orbital
ellipse. (a) WCK 1Aa,Ab = HIP 5336, (b) YSC 129 = HIP 32040, (c) LSC 45 = HIP 36387, 
(d) DSG 7Aa,Ab = HIP 81023, 
and (e) DSG 8 = HIP 117415.
}
\end{figure}

\begin{figure}[tb]
\epsscale{1.0}
\plotone{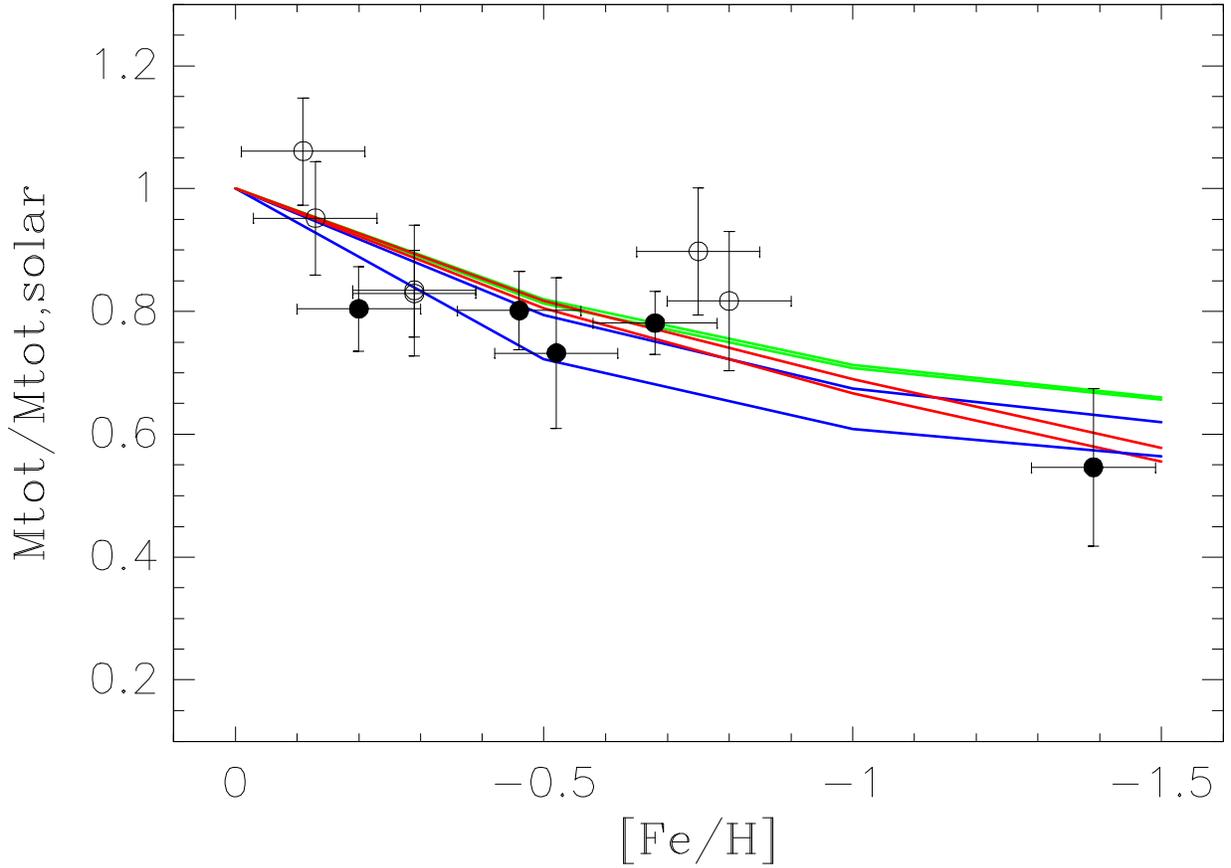}
\caption{
The observed dependence of total mass as a function of iron abundance
at fixed $B-V$ color. 
This plot is a revision of that presented
in Horch \ea (2015a) where the blue, green, and red curves represent 
values derived from stellar models as discussed in the text for main-sequence binaries of 
composite spectral type F, G, and K, respectively. There are two curves 
for each spectral type, for ages of 1 and 4 Gyrs. The open circles represent
systems that appeared in Horch \ea 2015a with revised placement based on
{\it Gaia} DR2 parallaxes and/or more recent orbit calculations whenever available, and the
filled circles are the systems shown in Table 5 and Figure 6 (two of which 
also appeared in the earlier paper).}
\end{figure}

\begin{figure}[tb]
\plotone{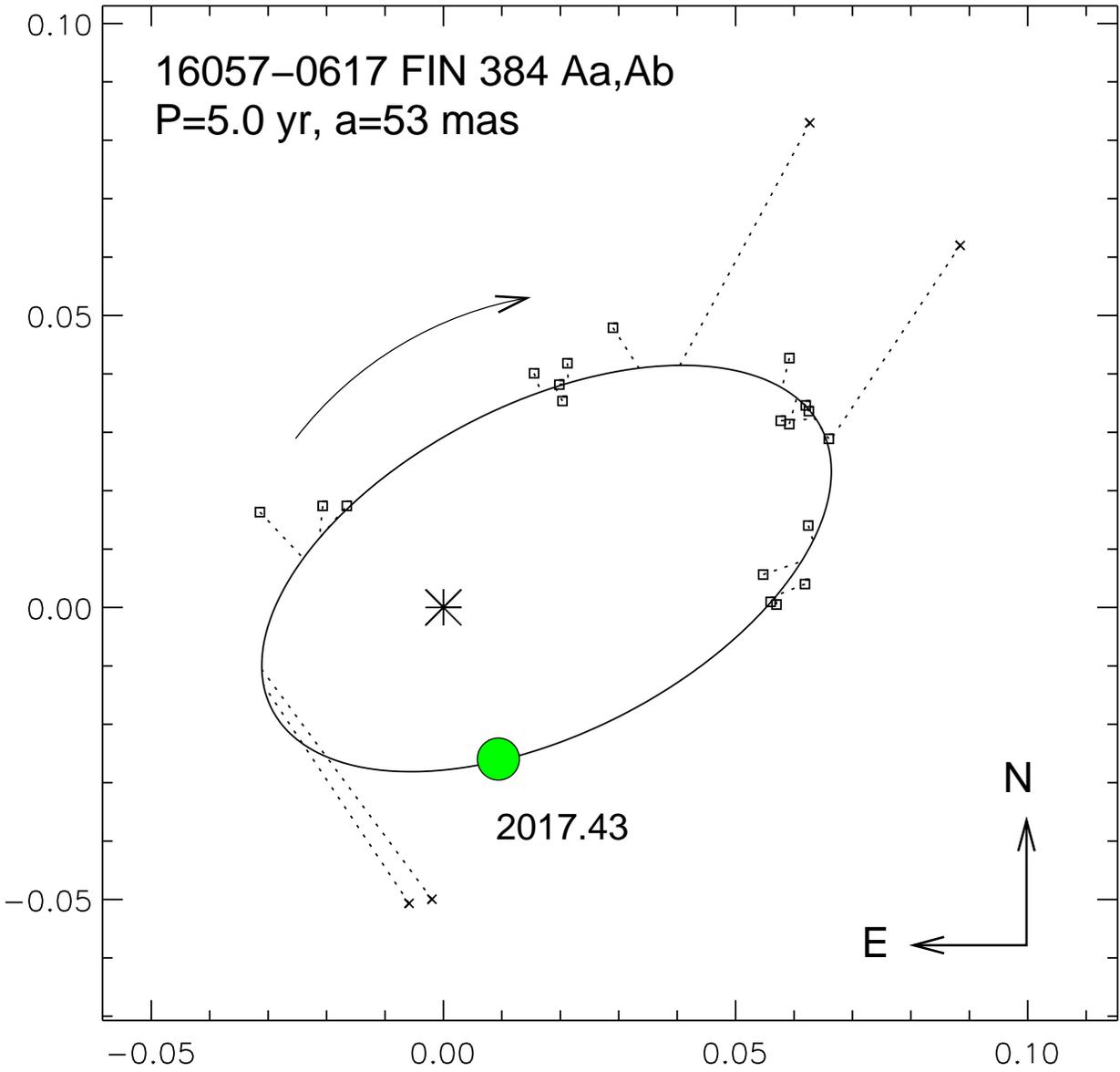}
\caption{
The inner orbit of the subsystem Aa,Ab in
the triple star HIP~78849 (BU 914 AB and FIN 384 Aa,Ab). The DSSI
measure is marked by the green circle, on the previously
unobserved part of the orbit. Four deviant measures (crosses) come from
micrometer observations and speckle interferometry, illustrating
the difficulty of this object. The outer orbit with a period of 311
yr is also known, and it is not coplanar with the inner orbit. 
}
\end{figure}

\end{document}